\documentclass[useAMS,usenatbib,usegraphicx]{mn2e}
\usepackage{bm}
\usepackage{times}

\newcommand{\ZOsun}{\mathrm{Z}_{\mathrm{O\sun}}}
\newcommand{\ZO}{Z_\mathrm{O}}
\newcommand{\Msun}{\mathrm{M}_{\sun}}
\newcommand{\nH}{n_{\mathrm{H}}}
\def\sdss{SDSS J1048+4637}

\title[Effects of grain shattering on extinction curves]
{Effects of grain shattering by turbulence on extinction curves
in starburst galaxies}
\author[Hirashita et al.]{Hiroyuki Hirashita$^1$\thanks{E-mail:
    hirashita@asiaa.sinica.edu.tw},
Takaya Nozawa$^2$, Huirong Yan$^{3}$ and
Takashi Kozasa$^4$
\\
$^1$Institute of Astronomy and Astrophysics, Academia Sinica,
P.O. Box 23-141, Taipei 10617, Taiwan\\
$^2$Institute for the Physics and Mathematics of the Universe,
University of Tokyo, Kashiwa 277-8568, Japan\\
$^3$University of Arizona, LPL, Steward Observatory and
Physics Department, 933 N
Cherry Avenue, Tucson, AZ 85721, USA\\
$^4$Department of Cosmosciences, Graduate
     School of Science, Hokkaido University, Sapporo
     060-0810, Japan
}
\date{2010 January 15}

\pagerange{\pageref{firstpage}--\pageref{lastpage}} \pubyear{2010}

\begin{document}
\label{firstpage}
\maketitle

\begin{abstract}
Dust grains can be efficiently accelerated and shattered in
warm ionized medium (WIM) because of the turbulent motion.
This effect is enhanced in starburst galaxies, where
{gas is ionized and} turbulence is sustained by
massive stars. Moreover, dust production by Type II
supernovae (SNe II) can be
efficient in starburst galaxies. In this paper, we examine the
effect of shattering in WIM on the dust grains produced
by SNe II. We find that although the
grains ejected from SNe~II are expected to be biased to
large sizes ($a\ga 0.1~\micron$, where $a$ is the grain radius)
because of  {the shock destruction in supernova
remnants},
the shattering in WIM is efficient enough in $\sim 5$ Myr
to produce small grains if the metallicity is nearly solar
or more.
The production of small grains by
shattering steepens the extinction curve. Thus,
steepening of extinction curves by shattering should
always be taken into account for the system where the
metallicity is solar and the starburst age is typically
larger than 5 Myr. These conditions
may be satisfied not only in nearby starbursts
but also in high redshift ($z>5$) quasars.
\end{abstract}


\section{Introduction}

Type II supernovae (SNe II) are {considered to be
one of the} grain production
sources in the Universe
\citep[e.g.][]{kozasa89,todini01,nozawa03}.
The significance of SNe II in the grain production is
enhanced if the cosmic age is so young (typically at
redshift $z>5$) that low-mass stars, i.e.\ asymptotic
giant branch (AGB) stars and Type Ia supernovae, cannot
contribute significantly to the dust formation
(\citealt{dwek07}; but see \citealt{valiante09}), or
if the current starburst is strong enough.
For some nearby blue compact dwarf galaxies (BCDs),
the latter condition may be satisfied
\citep{hirashita04,takeuchi05}. Thus, the
dust production by SNe II is tested by high $z$ objects
and nearby starbursts.

In galaxies where active star formation (starburst) is
occurring, dust grains are not only produced and
ejected from stars but also processed in the
interstellar medium (ISM). In such star-forming
galaxies, it is expected that the supernova (SN) rate
is enhanced, which leads to an efficient destruction
of large grains with
$a\ga 0.1~\micron$, where $a$ is the grain radius,
by SN shocks
\citep*[e.g.][]{jones94,jones96}. Also a large amount
of ionizing photons are
supplied and H\,\textsc{ii} regions develop. Indeed
giant H\,\textsc{ii} regions with sizes of $\ga 100$ pc
are found in nearby galaxies \citep{kennicutt84}.
Such large H\,\textsc{ii} regions can also be
expected theoretically with a starburst
\citep{hirashita04}. Moreover, the size is larger
than the typical expansion radius of SN
shells when the dust condensed in SNe II is finally
supplied to the ISM \citep[hereafter N07]{nozawa07}.
Thus, it is
reasonable to consider that
the dust ejected from SNe II is supplied to the ionized
regions in starburst galaxies.

\citet[hereafter HY09]{hirashita09} show that
shattering occurs efficiently in warm ionized medium
(WIM), where grains
are efficiently accelerated by
magnetohydrodynamic (MHD) turbulence
\citep*{yan03,yan04}.
{The same mechanism is also expected
to work in the ionized regions of starburst galaxies.}
Indeed, turbulence is ubiquitous in the ISM,
and the collective effects of OB stellar winds and
supernovae (SNe) can play an important role in
sustaining turbulence \citep[e.g.][]{elmegreen04}.
{Therefore, it is probable that the grains
ejected from SNe II into the ionized regions are
efficiently shattered by turbulence.}

Since large grains with $a\ga 0.1~\micron$ are
suggested to be injected into the ISM from SNe II
selectively due to the destruction in hot plasma
produced by {the reverse and forward shocks in
supernova remnants (SNRs)}
(N07)\footnote{{We call
this destruction process `destruction in SNRs'.}},
shattering in WIM is important for the production of
small grains.
If a significant amount of small grains
are produced, optical--ultraviolet (UV) grain opacity
is enhanced and the slope of the extinction curve
becomes steep. In particular, the extinction curves
of young starbursts are used to constrain the
composition and size distribution of grains formed
in SNe II \citep{maiolino04,hirashita05}. Thus, it
is important to quantify the effect of grain shattering
on the extinction curve.

Also in the observational context, production of
small grains in starburst environment is worth
investigating. The modification of grain size
distribution should have an impact on the dust
extinction and emission properties
\citep[e.g.][]{dopita05}. By studying the spectral
energy distributions (SEDs) of dust and stars of
some actively star-forming dwarf galaxies,
\citet{galliano05} show that the grain size
distribution is biased to small sizes
($\sim$ a few nm).
The extinction curves of
starburst galaxies in general show a
significant reddening in the optical--UV range
\citep{calzetti01}, 
indicating that there should be some contribution
from small grains. \citet{galliano05}
also suggest
that their results are consistent with shattering
and erosion of ISM grains by SN shocks
(\citealt{jones94,jones96}; see also
\citealt{borkowski95}). Both shock and turbulence
are efficient drivers of the relative motion between
grains, since the grain acceleration occurs in a
way strongly dependent on the grain size
{\citep{shull77,mckee87,jones96,yan04}}.%
\footnote{Galactic-scale
bulk motions such as collective
outflow from stellar feedbacks are not efficient
in producing the relative motions among grains.}
However, at present,
it is not
known which of the two drivers are
more important.
Thus, in this paper, as a first necessary step to
fully understand the possible mechanisms of
small grain production, we focus on turbulence
as a possible driver of grain shattering.

This paper is organized as follows. We explain the
method used in this paper in
Section \ref{sec:method}, and describe
some basic results of our calculations in
Section \ref{sec:result}. We discuss the results
and mention some observational implications in
Section \ref{sec:discussion}.
Finally, Section \ref{sec:conclusion} gives the conclusion.

\section{Method}\label{sec:method}

We consider a young starburst, where dust is
predominantly supplied by SNe II. We then calculate the
modification of grain size distribution by shattering in
WIM. Finally, we examine if the shattering effect is
apparent in the extinction curve or not.  In this section,
we first explain the initial grain size distribution in the
ejection from SNe II (Sections
\ref{subsec:initial} and \ref{subsec:norm}).
Next we review
the treatment of shattering which was
 {adopted from \citet{jones94,jones96}} in
the previous paper (Section \ref{subsec:shatter}). We
use the grain velocities calculated by a MHD
turbulence model to obtain the relative grain
velocities in
shattering (Section \ref{subsec:vel}). We also describe
the calculation method of extinction curve
(Section \ref{subsec:extinc}).
Throughout this
paper, grains are assumed to be
spherical with radius $a$.

\subsection{Initial grain size distribution}
\label{subsec:initial}

The size distribution of grains ejected from SNe II into
the ISM (WIM) is adopted from N07. This size distribution
is used as the initial condition for the calculation of
shattering in WIM. N07 have shown that the size
distribution of dust formed in the ejecta is largely
modified by the destruction by sputtering in ionized gas
heated by the reverse and forward shocks. Not only N07
but also \citet{bianchi07} treat the effect of shock
destruction in SNe II. N07 consider some aspects which
\citet{bianchi07} did not take into account: N07 solve the
motion of dust grains
by taking into account the gas drag and treat the
destruction of dust in 
the radiative phase as well as in the non-radiative phase of
SNRs. Thus, we adopt N07's
results for the size distribution
of grains ejected from SNe II into the ISM, although
qualitative behaviour of our results are common
even if we adopt the size distribution of \citet{bianchi07}.
{
Also we should mention that both N07 and \citet{bianchi07}
neglect the effects of dust electrical charge and the
effects of magnetic fields on grain kinematics. These physical
processes, which should be quantified in future work, are
further discussed in Section \ref{subsec:remark}.
}

Here we briefly summarize the calculation of N07. N07
started from the grain size distribution calculated by
\citet{nozawa03}, who treated dust nucleation and
growth in SNe II based on
the SN model of \citet{umeda02}. Then,
N07 took
into account the dust destruction by {kinetic and thermal
sputtering in hot gas swept up by the reverse and forward
shocks}
after the interaction of the SN ejecta with the ambient
ISM. Thus, the grain size
distribution calculated by N07 is regarded as that
ejected from SNe II to the ISM.

N07 treated two extreme cases for the mixing of
elements in a SN II: one is the unmixed case in which
the original onion-like structure of elements is
preserved, and the other is the mixed case
in which the elements are uniformly mixed within
the helium core.
In this paper, we adopt the unmixed
case, since the extinction features of carbon and
silicon, which are major grain components in the
unmixed case, are consistent with observations
\citep{hirashita05,kawara10}.
{Even if the mixed case is adopted, the grain size
distribution is similarly biased to large grain sizes,
so that the behaviour of the extinction curve calculated
later is expected to be similar.}

As a representative progenitor mass, we adopt
20 $\Msun$, following N07. The formed grain species
are C, Si, SiO$_2$, Fe, FeS, Al$_2$O$_3$, MgO,
MgSiO$_3$, and Mg$_2$SiO$_4$. According to
their calculation, {small grains with
$a\la 0.02~\micron$ are trapped in the
shocked region because the deceleration
rate of a grain by the gas drag is inversely
proportional to its size \citep[e.g.][]{nozawa06}.
Thus, these small grains
are efficiently destroyed by
thermal sputtering} if the
ambient hydrogen number density,
$\nH$, is larger than 0.1 cm$^{-3}$. Moreover,
the destruction efficiency depends sensitively on $\nH$.
With $\nH$ as large as 10 cm$^{-3}$, only a few percent
of grains survive and the grain size distribution is
strongly biased to large ($a\ga 0.1~\micron$) radii.
{It is interesting to point out that \citet*{slavin04}
derived similar grain radii for the dynamical decoupling
of grains from the interstellar gas, although they include
magnetic fields in their calculation (see also
Section \ref{subsec:remark}). The following results on the
effects of grain shattering by turbulence in  WIM are
not largely
affected as long as the initial size distribution is
biased to large  ($a\simeq 0.1~\micron$)
grains.}

\subsection{Normalization of grain size distribution}
\label{subsec:norm}

The total grain mass density integrated for all the
size range, $\rho_\mathrm{dust}$, is normalized to the
gas mass density, $1.4\nH m_\mathrm{H}$, to obtain the
dust-to-gas ratio,
$\mathcal{D}\equiv\rho_\mathrm{dust}/(1.4\nH m_\mathrm{H})$,
where $m_\mathrm{H}$ is
the mass of hydrogen atom and the factor 1.4 is the
correction for the species other than hydrogen.
Since dust grains are composed of metals, it is useful
to label the dust abundance in terms of metallicity.
We parameterize
the dust abundance by the oxygen abundance, because
oxygen is one of the main metals produced by SNe~II
and oxygen emission lines in the optical are often
used to estimate the gas-phase metal abundance.
The oxygen mass produced by a SN II of
20 $\Msun$ progenitor is
$m_\mathrm{O}=1.58~\Msun$ \citep{umeda02}.
The dust mass after the {destruction in SNRs}
($m_\mathrm{d}$)
is listed for each ambient density in
Table \ref{tab:model}. The solar oxygen abundance is
assumed to be
$\ZOsun =9.7\times 10^{-3}$ in mass ratio \citep{anders89}.
(The oxygen abundance is denoted as $\ZO$ and
is simply called metallicity in this paper.)
Therefore, in the solar metallicity case, we
assume that the dust-to-gas ratio is
$\mathcal{D}_0=\ZOsun m_\mathrm{d}/m_\mathrm{O}$,
which is also listed in Table \ref{tab:model}. The
dust-to-gas ratio is assumed to be proportional to
the metallicity:
$\mathcal{D}=(\ZO /\ZOsun )\mathcal{D}_0$. This is
equivalent to the assumption that both dust and
oxygen are predominantly supplied from SNe II.

\begin{table}
\centering
\begin{minipage}{55mm}
\caption{Characteristic quantities for SN II dust
production for various ambient densities $\nH$.}
\label{tab:model}
    \begin{tabular}{ccc}
     \hline
     $\nH$ (cm$^{-3}$) & $m_\mathrm{d}$ ($\Msun$) &
     $\mathcal{D}_0$ \\ \hline 
     0.1 & 0.35 & $2.2\times 10^{-3}$
             \\
     1   & 0.14 & $8.7\times 10^{-4}$
             \\
     10  & 0.033 & $2.0\times 10^{-4}$
             \\
      \hline
    \end{tabular}

\textit{Note.} $m_\mathrm{d}$ is the dust mass ejected per
SN II (a progenitor mass of 20 $\Msun$ is assumed), and
$\mathcal{D}_0$ is the dust-to-gas ratio at
the solar metallicity (oxygen abundance) $\ZOsun$.
\end{minipage}
\end{table}

\subsection{Shattering}\label{subsec:shatter}

Dust grains suffer shattering if the relative velocity
between grains is larger than 2.7 and 1.2 km s$^{-1}$
for silicate and graphite, respectively
\citep{jones96}. As shown by \citet{yan04}, these
velocities can be achieved in magnetized and turbulent
ISM. In particular, grains are efficiently accelerated
in WIM, since damping is weak \citep{yan04}. The
parameters adopted for MHD
turbulence and the grain velocities obtained are
described in Section \ref{subsec:vel}.
The time evolution of grain size distribution by
shattering is calculated by adopting the formulation
of  {\citet{jones94,jones96}}.
We briefly review the calculation
method. The details are described in HY09.

We solve the shattering equation discretized for the
grain size. Although nine grain species are treated
here (Section \ref{subsec:initial}), the material
properties needed for the calculation of shattering are
not necessarily available for all the species. Thus, we
divide the grains into two groups: one is carbonaceous
dust and the other is all the other species of dust,
and apply the relevant material quantities of graphite
and silicate, respectively. In fact, as shown later,
the mass and opacity of the latter group is dominated
by Si. The validity of this approximation that all the
species other than carbonaceous dust are treated as
silicate (called one-species method) is examined in
comparison with another extreme
approximation (individual-species method) in the
Appendix \ref{app:individual}.
Because of the lack of the experimental data for
Si, we assume that Si (and also the other `silicate'
species) can be treated as silicate in
shattering because of similar
hardness.\footnote{Among the materials whose shattering
properties are available in Table 1 of \citet{jones96}
(i.e.\ silicate, SiC, ice, iron, and diamond), silicate
is expected to have the nearest atomic binding energy.
{As shown by \citet{serra08},
who derived the previously unknown shattering
properties of hydrogenated amorphous carbon,
it is not impossible to estimate relevant quantities, but
some guiding quantities from other materials is still
necessary even in their case.
Thus, here we simply
adopt the material quantities of silicate for Si.}}
Since the cratering volume in a grain--grain
collision is approximately proportional to
$1/P_\mathrm{cr}$, where $P_\mathrm{cr}$ is the
critical shock pressure
for shattering \citep{jones96}, the shattering time-scale
is roughly proportional to $P_\mathrm{cr}$. Thus,
if $P_\mathrm{cr}$ is obtained for appropriate
materials (especially Si) in some future experiment,
our results can easily be scaled. The material properties
of silicate and graphite are taken from
\citet{jones96} and summarized in HY09.

The number density of grains whose radii are between
$a$ and $a+\mathrm{d}a$ is denoted as
$n(a)\,\mathrm{d}a$, where the entire range of $a$ is
from $a_\mathrm{min}$ to
$a_\mathrm{max}$. To ensure the conservation of
the total mass of grains, it is numerically convenient to
consider the distribution function of grain mass instead
of grain size. We denote the number density of grains
whose masses are
between $m$ and $m+\mathrm{d}m$ as
$\tilde{n}(m)\,\mathrm{d}m$.
The two distribution functions are related as
$n(a)\,\mathrm{d}a=\tilde{n}(m)\,\mathrm{d}m$
and $m=(4\pi /3)a^3\rho_\mathrm{gr}$, where
$\rho_\mathrm{gr}$
is the grain material density (3.3 and 2.2 g cm$^{-3}$
for silicate and graphite, respectively).

For numerical calculation, we consider $N$ discrete
bins for the grain radius. The grain radius in the
$i$-th ($i=1,\,\cdots ,\, N$) bin is between
$a_{i-1}^\mathrm{(b)}$ and $a_i^\mathrm{(b)}$, where
$a_{i}^\mathrm{(b)}=a_{i-1}^\mathrm{(b)}\delta$,
$a_0^\mathrm{(b)}=a_\mathrm{min}$, and
$a_N^\mathrm{(b)}=a_\mathrm{max}$ (i.e.\
$\log\delta$ specifies the width of a logarithmic bin:
$\log\delta =(1/N)\log (a_\mathrm{max}/a_\mathrm{min})$).
We represent the grain radius and mass in the $i$-th bin
with
$a_i\equiv (a_{i-1}^\mathrm{(b)}+a_i^\mathrm{(b)})/2$ and
$m_i\equiv (4\pi /3)a_i^3\rho_\mathrm{gr}$. The boundary
of the mass bin is defined as
$m_i^\mathrm{(b)}\equiv (4\pi /3)[a_i^\mathrm{(b)}]^3
\rho_\mathrm{gr}$.
Giving $a_\mathrm{min}$, $a_\mathrm{max}$, and $N$,
all bins can be set. A grain in the $i$-th bin
is called ``grain $i$''. We adopt
$a_\mathrm{min}=3\times 10^{-4}~\micron$ (3 \AA) and
$a_\mathrm{max}=3~\micron$ to cover the entire grain
size range in N07, and $N=32$.
{We have confirmed that the results are not
altered even if $N$ is doubled.}

The mass density of grains contained in the $i$-th bin,
$\tilde{\rho}_i$, is defined as
\begin{eqnarray}
\tilde{\rho}_i\equiv m_i\tilde{n}(m_i)(m_i^\mathrm{(b)}-
m_{i-1}^\mathrm{(b)})\, .
\end{eqnarray}
Note that $\tilde{\rho}_i=\rho_i\delta_i$ in the expression
of \citet{jones94,jones96}. The time evolution of
$\tilde{\rho}_i$ by
shattering can be written as
\begin{eqnarray}
\left[\frac{\mathrm{d}\tilde{\rho}_i}{\mathrm{d}t}
\right]_\mathrm{shat}\hspace{-2mm}
& =\hspace{-2mm} &
-m_i\tilde{\rho}_i
\sum_{k=1}^{N}\alpha_{ki}\tilde{\rho}_k+
\sum_{j=1}^{N}\sum_{k=1}^N\alpha_{kj}\tilde{\rho}_k
\tilde{\rho}_jm_\mathrm{shat}^{kj}(i)\, , \nonumber\\
& &
\label{eq:time_ev}
\end{eqnarray}
\begin{eqnarray}
\alpha_{ki}=\left\{
\begin{array}{ll}
{\displaystyle \frac{\sigma_{ki}v_{ki}}{m_im_k}} &
\mbox{if $v_{ki}>v_\mathrm{shat}$,} \\
0 & \mbox{otherwise,}
\end{array}
\right.
\end{eqnarray}
where $m_\mathrm{shat}^{kj}(i)$ is the total mass of the
shattered fragments of a grain $k$ that enter the $i$-th
bin in the collision between grains $k$ and $j$,
$\sigma_{ki}$ and $v_{ki}$ are, respectively, the
grain--grain
collision cross section and the relative collision speed
between grains $k$ and $i$, and $v_\mathrm{shat}$ is the
velocity threshold for shattering to occur.
For the cross section, we apply
$\sigma_{ki}=\pi (a_k+a_i)^2$.

The grain velocities given by \citet{yan04} are typical
velocity dispersions. In order to take into account the
directional information, {we follow the method in
\citet{jones94}:} we divide each time step into 4
small steps, and we apply
$v_{ik}=v_i+v_k$, $|v_i-v_k|$, $v_i$, and $v_k$ in each
small step, where $v_i$ and $v_k$
are the velocities of grains $i$ and $k$, respectively
(see Section \ref{subsec:vel}).
Note that the mass of the
shattered fragment $m_\mathrm{shat}^{kj}(i)$ depends on
$v_{kj}$ as described in  {\citet{jones96}}.
Briefly, the total fragment mass is determined by the
shocked mass in the collision and the fragments are
distributed with a grain size distribution
$\propto a^{-\alpha_\mathrm{f}}$ with
$\alpha_\mathrm{f}=3.3$ unless otherwise stated
(see HY09 for the size range of the fragments).

{
For the shattering duration, several Myr may be
appropriate, since it is a typical lifetime of ionizing
stars with a mass $\ga 20~\Msun$ (lifetime $<10$ Myr)
\citep{bressan93,inoue00}. HY09 also indicate that the
small ($a\la 0.01~\micron$) grains depleted
by coagulation in dense clouds are recovered if
shattering in WIM lasts for 3--5 Myr, which
justifies the necessity of shattering in WIM for
mega-years}.
We also examine a longer
time-scale, 10 Myr, to investigate a starburst environment
where an intense star formation occurs continuously. Such
a situation may be realized in extragalactic giant
ionized regions \citep{hirashita04,hunt09}.
In summary, we examine $t=3$, 5, and 10 Myr, where $t$ is
the elapsed time of shattering.

\subsection{Grain velocity}\label{subsec:vel}

The grain velocity as a function of grain radius $a$ in
the presence of interstellar MHD turbulence is calculated
by the method
described in \citet{yan04}. They considered the grain
acceleration by hydrodrag and gyroresonance
and calculated the grain velocities achieved in various
phases of ISM. Among the ISM phases, we focus on WIM to
investigate the possibility of efficient shattering in
actively star-forming environments.

We adopt three cases for the hydrogen number density
of WIM ($\nH =0.1$, 1, and 10 cm$^{-3}$), since N07
applied these densities for the
ambient medium. {For WIM, a density of
$\nH\sim 0.1$--1 cm$^{-3}$ is usually considered
\citep{mckee77}, but we also examine a density
as high as $\nH\sim 10$ cm$^{-3}$ for
young H \textsc{ii} regions around massive stars as
observed in starburst environments
\citep{hunt09}. Embedded starbursts may also have
such dense ionized regions.}
We adopt gas temperature
$T=8000$ K, electron number density
$n_\mathrm{e}=\nH$, Alfv\'{e}n speed
$V_\mathrm{A}=20$ km s$^{-1}$
and injection scale of
the turbulence $L=100$ pc, following \citet{yan04}.
The effect of the injection scale is minor to that of
the sound and Alfv\'{e}n velocities. Since both the
sound speed and the Alfv\'{e}n speed are fixed, the
plasma $\beta$ is constant in all cases.
The grain charge
is assumed to be the same as that in \citet{yan04},
who calculated it by assuming a typical Galactic
condition. Since we expect higher interstellar
radiation field and higher electron density for the
starburst environments, the absolute values for the
grain charge can be larger than those assumed here.
For grains with
$a\ga 0.1~\micron$, where most of
the grain mass is contained in our cases, the grain
velocity is
governed by the gyroresonance. The acceleration rate of
gyroresonance increases with the grain charge, but the
acceleration duration, the gaseous drag time, decreases
with the grain charge \citep{yan03}. As a result, the
acceleration efficiency of gyroresonance is insensitive
to the grain charge.

\begin{figure*}
\includegraphics[width=0.45\textwidth]{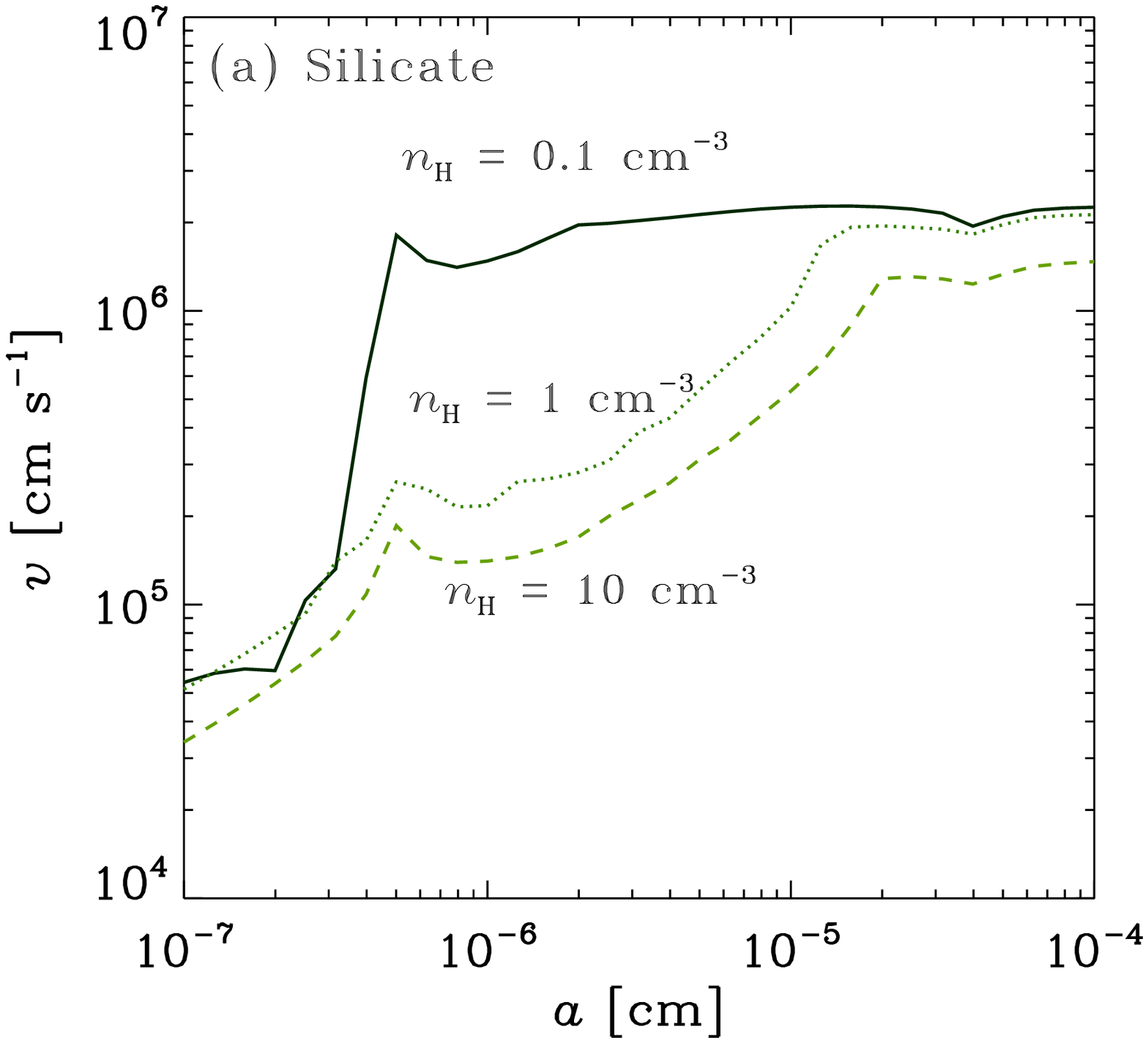}
\includegraphics[width=0.45\textwidth]{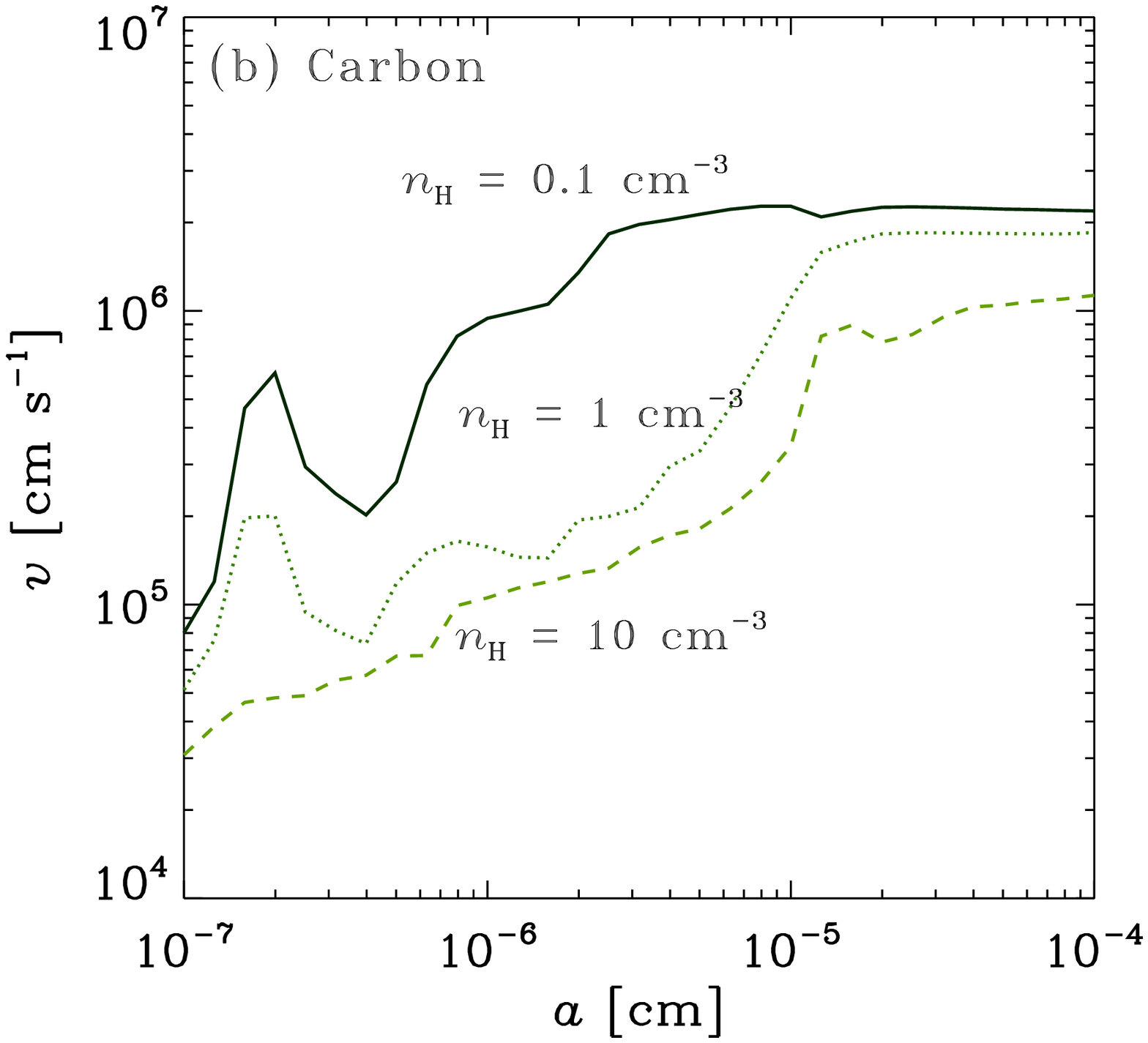}
 \caption{Grain velocities $v$ calculated from the
turbulence model as a function of grain radius $a$. Two
grain species, (a) silicate and (b) carbonaceous dust,
are shown. The solid,
dotted, and dashed lines indicate the velocities with
hydrogen number densities of
$\nH =0.1$, 1, and 10 cm$^{-3}$, respectively.}
 \label{fig:vel}
\end{figure*}

In Fig.\ \ref{fig:vel}, we show the grain velocities.
In general, larger grains tend to acquire larger
velocities because they are coupled with larger-scale
motions. For small grains, the motion is governed by
the gaseous drag, which has a linear dependence on the
grain charge \citep{yan04}. This is the reason for the
complex (non-monotonic) behaviour of
the grain velocity as a function of $a$ for small grains
($<0.1~\micron$).
We also observe that the grain velocity is not very
sensitive to $\nH$ for large ($a\ga 0.1~\micron$) grains,
whose shattering is important in this paper.

\subsection{Extinction curves}\label{subsec:extinc}

Extinction curves have been an effective tool to
examine the dust properties
{\citep[e.g.][]{mathis90}}.
For the calculation of extinction curves, we adopt
the same optical constants as those in
\citet{hirashita08}  {for the grain species formed
in SNe II (C, Si, SiO$_2$, Fe, FeS,
Al$_2$O$_3$, MgO, MgSiO$_3$, and
Mg$_2$SiO$_4$).
The grain properties of individual
species and the references for the optical constants
are listed in Table 1 of \citet{hirashita08}.}
By using those optical constants, we calculate the
absorption and scattering cross sections of
homogeneous spherical grains with various sizes
based on the Mie theory \citep{bohren83}. Then, the
grain extinction coefficient as a function of
wavelength is obtained by
weighting the cross sections with the grain size
distribution. The total extinction as a function of
wavelength $\lambda$, denoted as
$A_\lambda$, is calculated by
summing the contribution from all the species.

As stated in Section \ref{subsec:shatter}, we divide
the grain species into two groups in the calculation
of shattering: one is carbonaceous dust and the other
is silicate, which in fact contains all the species
other than carbonaceous dust. In the calculation of
extinction curves, the size distribution
of the silicate species is redistributed to each
component (Si, SiO$_2$, Fe, FeS, Al$_2$O$_3$, MgO,
MgSiO$_3$, and Mg$_2$SiO$_4$) in proportion to the
grain volume (i.e.\ the total mass of each component
divided by its material density) with a fixed shape of
the grain size distribution. In fact, Si is
dominated in the extinction curve, so that the
uncertainty coming from the above rough treatment
does not affect our conclusion
(Appendix \ref{app:individual}).

\section{RESULTS}\label{sec:result}

\subsection{Grain size distribution after shattering}

\begin{figure*}
\includegraphics[width=0.45\textwidth]{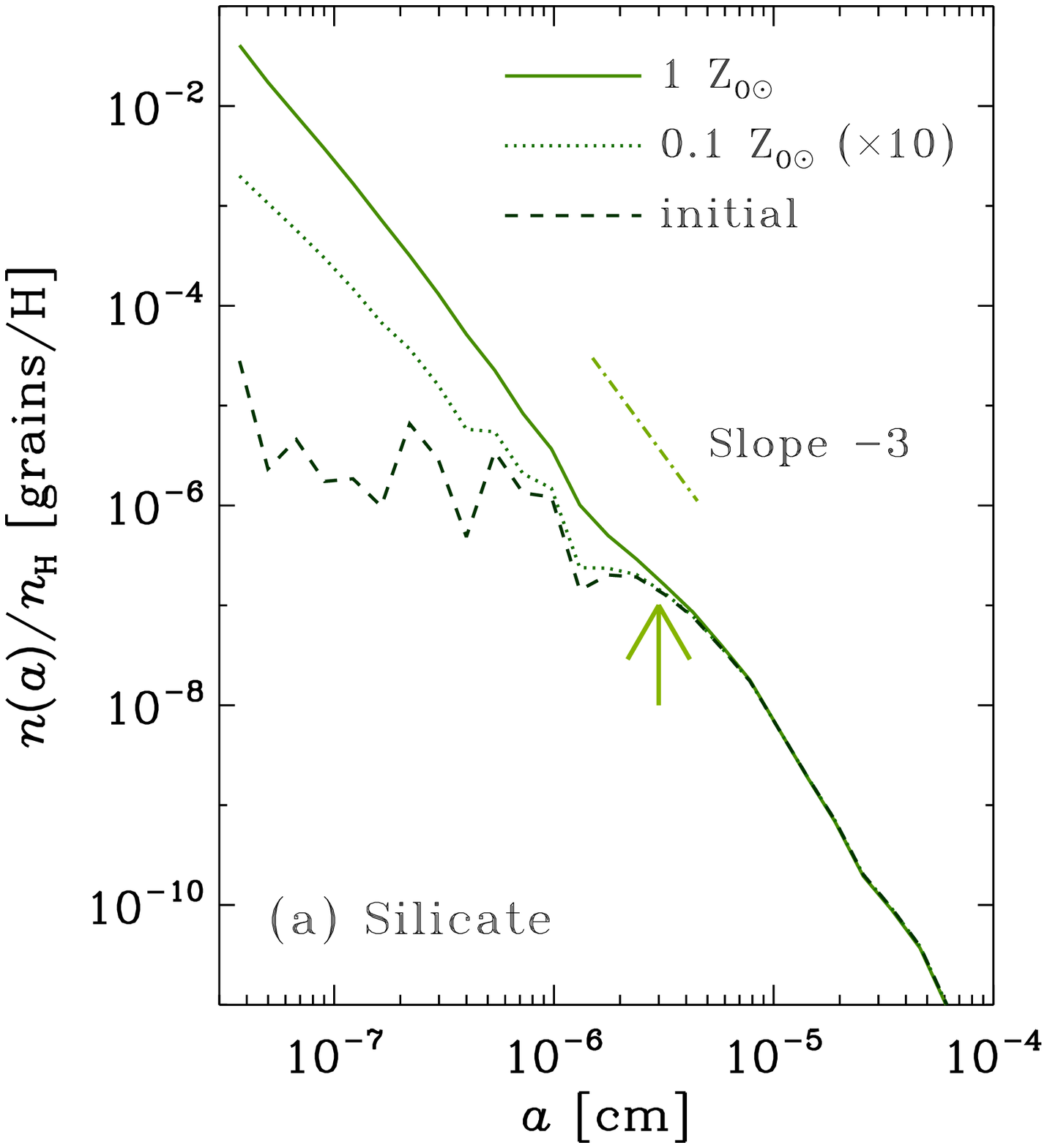}
\includegraphics[width=0.45\textwidth]{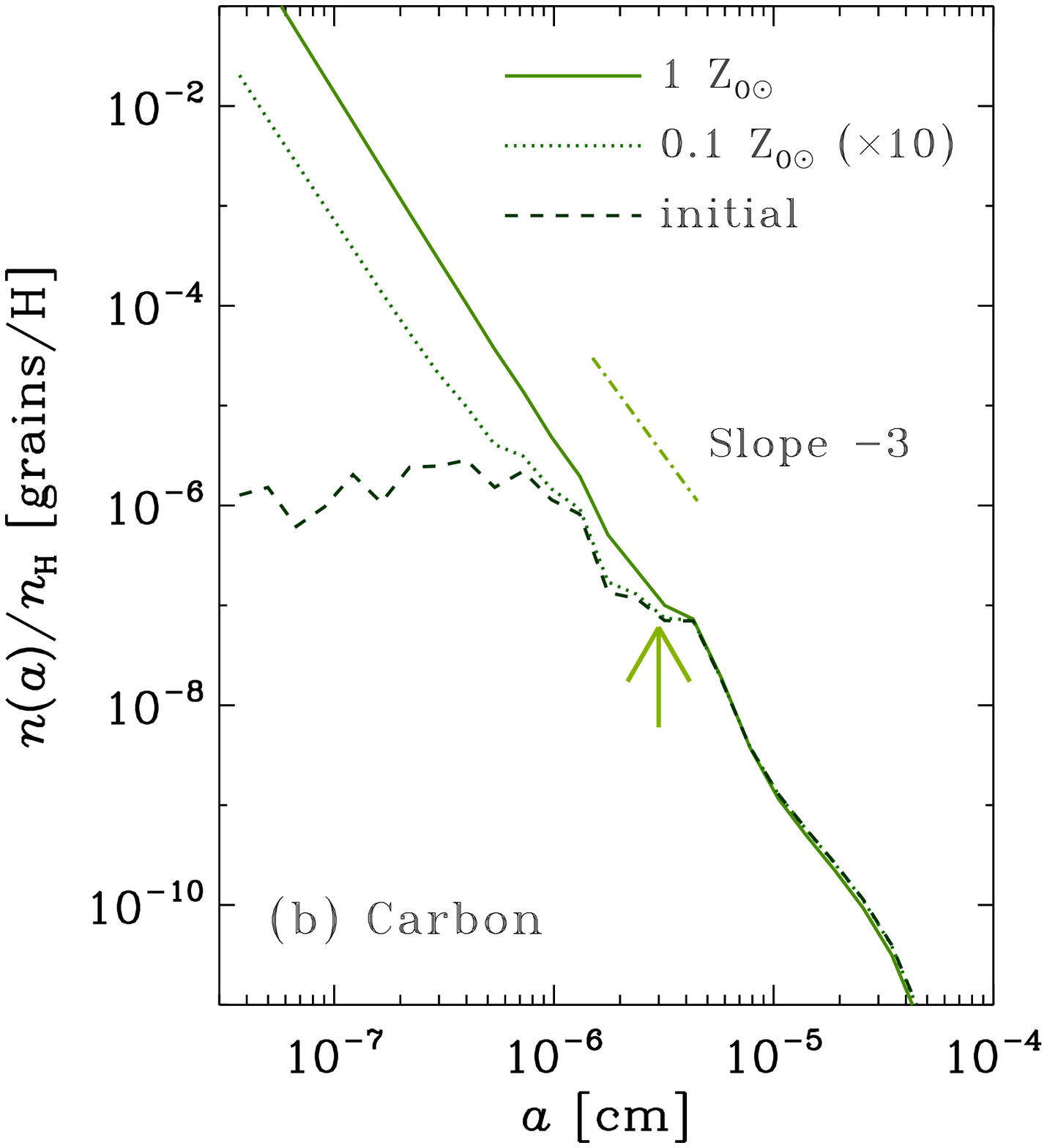}
 \caption{Grain size distributions per hydrogen atom.
The solid and dotted lines show the results at $t=5$ Myr
for metallicities of 1 $\ZOsun$ and 0.1 $\ZOsun$,
respectively. The hydrogen number density $\nH$ is assumed
to be 0.1 cm$^{-3}$. The dashed line presents the initial
grain size distribution before shattering.
Two grain species,
(a) silicate and (b) carbonaceous dust, are shown.
The case with 0.1 $\ZOsun$ is multiplied by 10 for the
convenience of presentation to offset the
10 times smaller dust abundance.
{The arrow is put at
$a=0.03~\micron$ as a rough representative size of the
grains contributing to the steepening of the UV
extinction curve.}}
 \label{fig:size_n0.1}
\end{figure*}

\begin{figure*}
\includegraphics[width=0.45\textwidth]{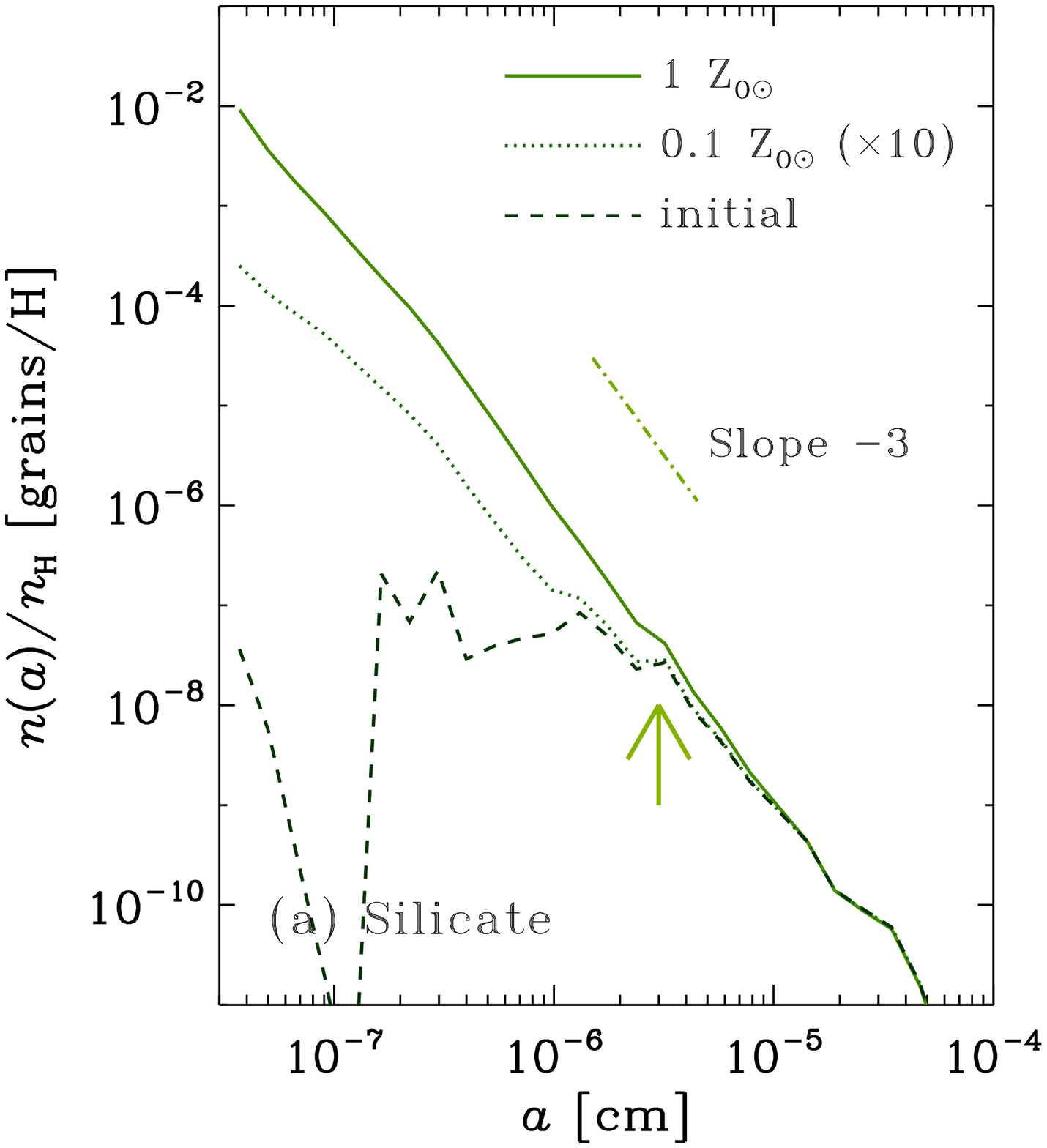}
\includegraphics[width=0.45\textwidth]{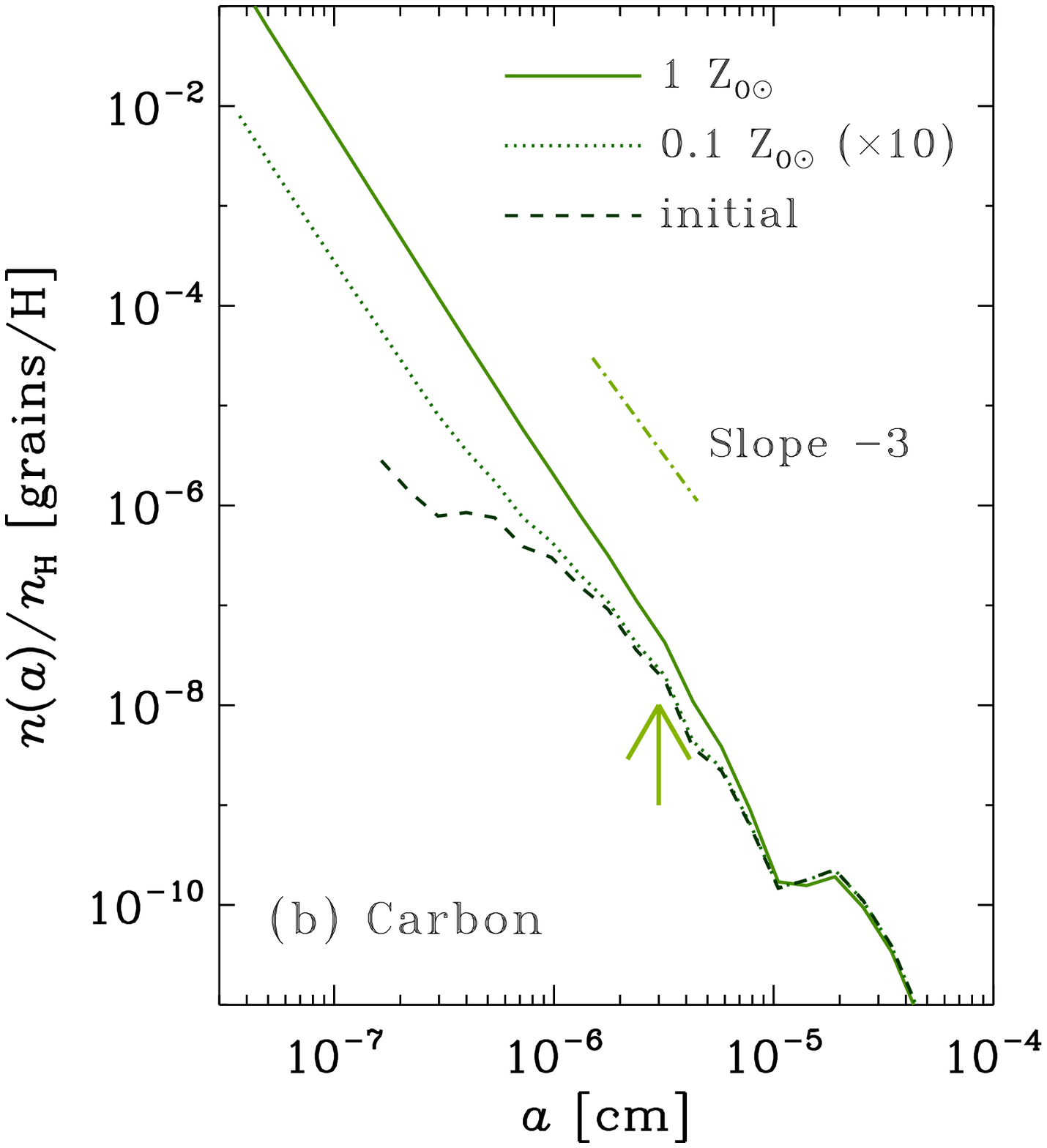}
 \caption{Same as Fig.\ \ref{fig:size_n0.1} but for
$\nH =1$ cm$^{-3}$.}
 \label{fig:size_n1}
\end{figure*}

\begin{figure*}
\includegraphics[width=0.45\textwidth]{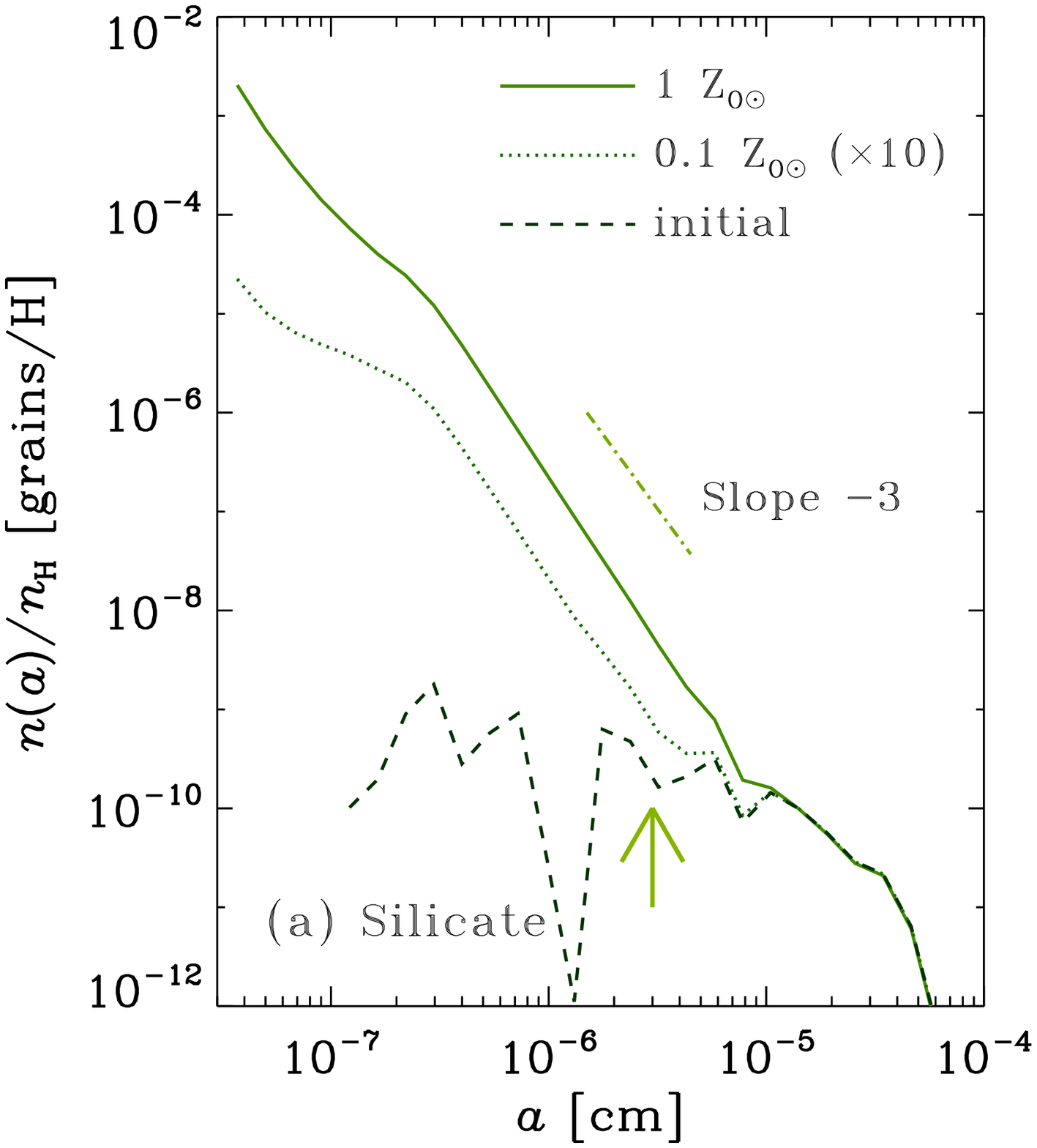}
\includegraphics[width=0.45\textwidth]{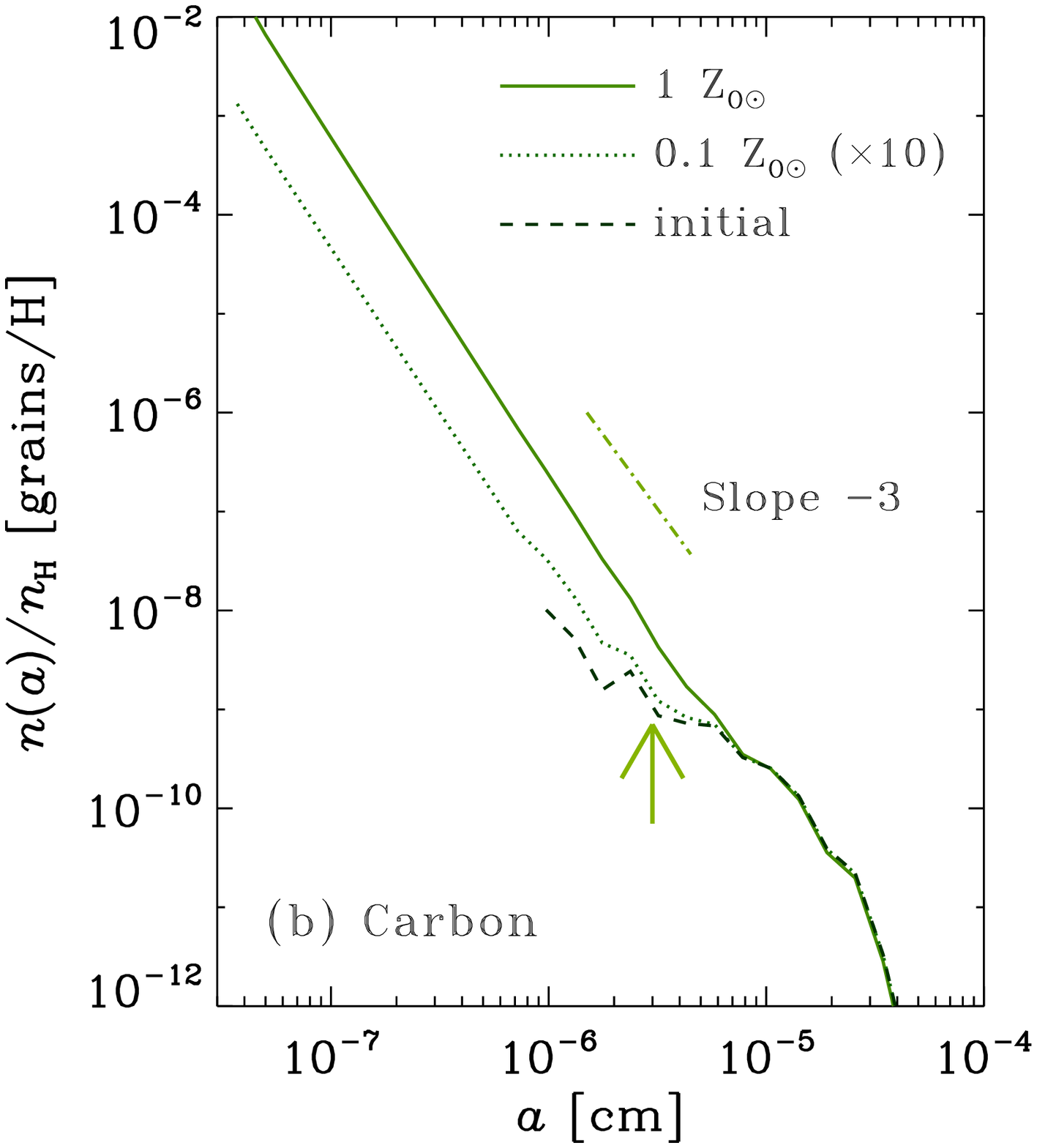}
 \caption{Same as Fig.\ \ref{fig:size_n0.1} but for
$\nH =10$ cm$^{-3}$.}
 \label{fig:size_n10}
\end{figure*}

The grain size distributions after shattering are
shown in Figs.\ \ref{fig:size_n0.1}--\ref{fig:size_n10}
for $\nH =0.1$, 1, and 10 cm$^{-3}$, respectively.
The grain size distribution is shown by $n(a)/\nH$.
We adopt $t=5$ Myr
as a typical time-scale on which WIM is sustained by
the radiation from {massive} stars
(Section \ref{subsec:shatter}). Two cases for the
metallicity, $\ZO =0.1$ and $1~\ZOsun$, are examined.

We observe that shattering really affects the grain size
distribution for all the densities. In particular, the
{abundance of small grains with
$a\la 0.1~\micron$ significantly increases after
shattering of a small portion of larger grains.}
If the metallicity is
1 $\ZOsun$, a continuous power-law-like size distribution
is realized for $a\la 0.1~\micron$.
Although the dust abundance is lower for
higher $\nH$
(Table \ref{tab:model}) because of more efficient
shock destruction in the SN remnant before the ejection
to the ISM (N07),
the grain--grain collision rate is enhanced in higher
$\nH$ environments.

The increase of grains with $a\la 0.1~\micron$ could
efficiently affect the UV and optical extinction curves.
This point is quantitatively addressed in
Section \ref{subsec:theor_extinc}. Large grains with
$a>0.1~\micron$ are marginally affected by shattering;
namely, shattering of a small fraction of large grains
can produce a large number of small grains. In the
case of HY09, on the other hand, grains
with $a>0.1~\micron$ are more shattered because
abundant small grains in
the MRN \citep{mathis77} grain size distribution,
which they assumed
as the initial condition, enhance the grain--grain
collision rate.

\subsection{Extinction curves}
\label{subsec:theor_extinc}

The extinction curve of grains ejected from SNe II
tends to be flat because small grains are efficiently
destroyed in SNRs without escaping into the ISM
\citep{hirashita08}. Here we investigate if the increase
of small grains by shattering
effectively steepens the extinction curves or not.

\begin{figure*}
\includegraphics[width=0.45\textwidth]{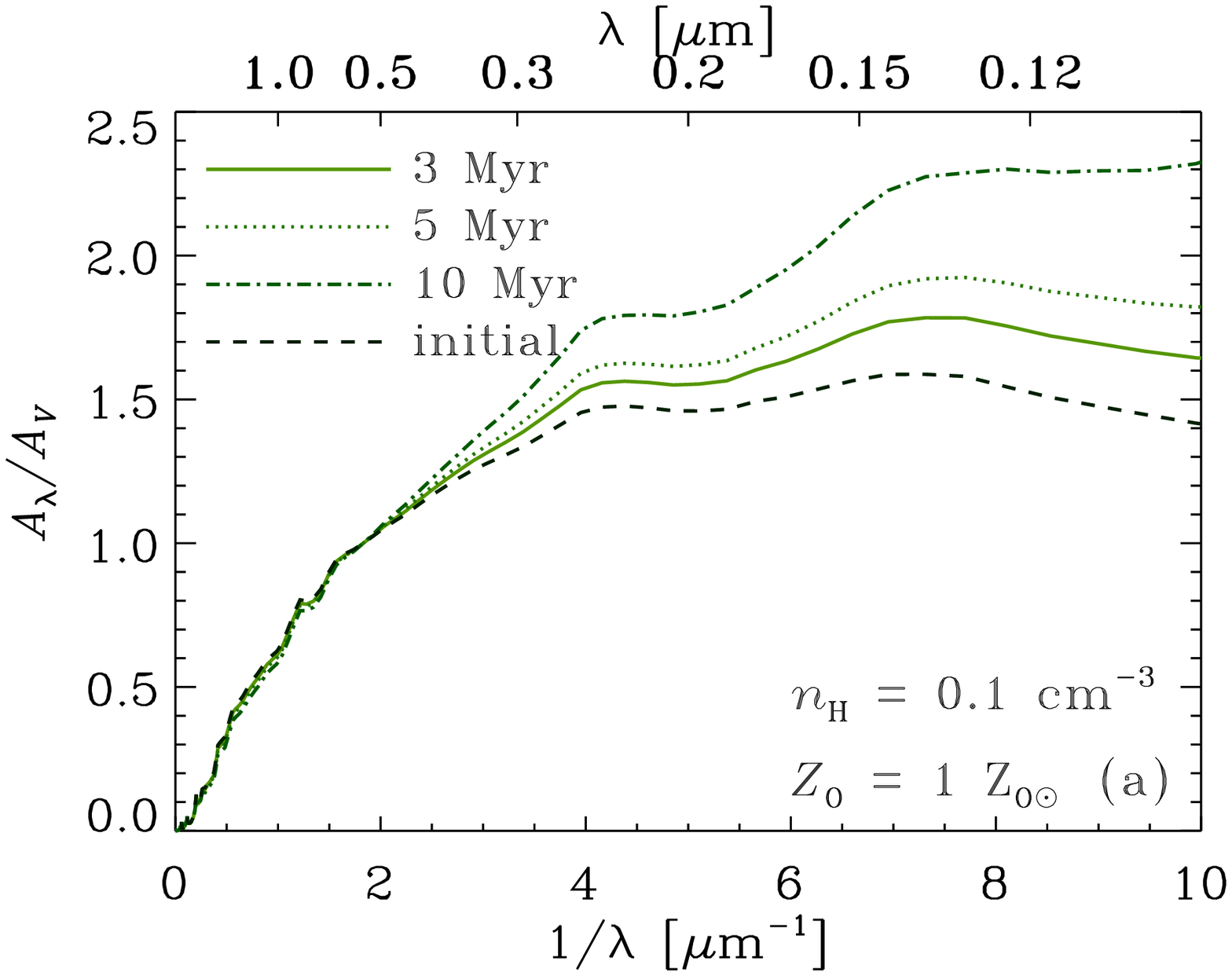}
\includegraphics[width=0.45\textwidth]{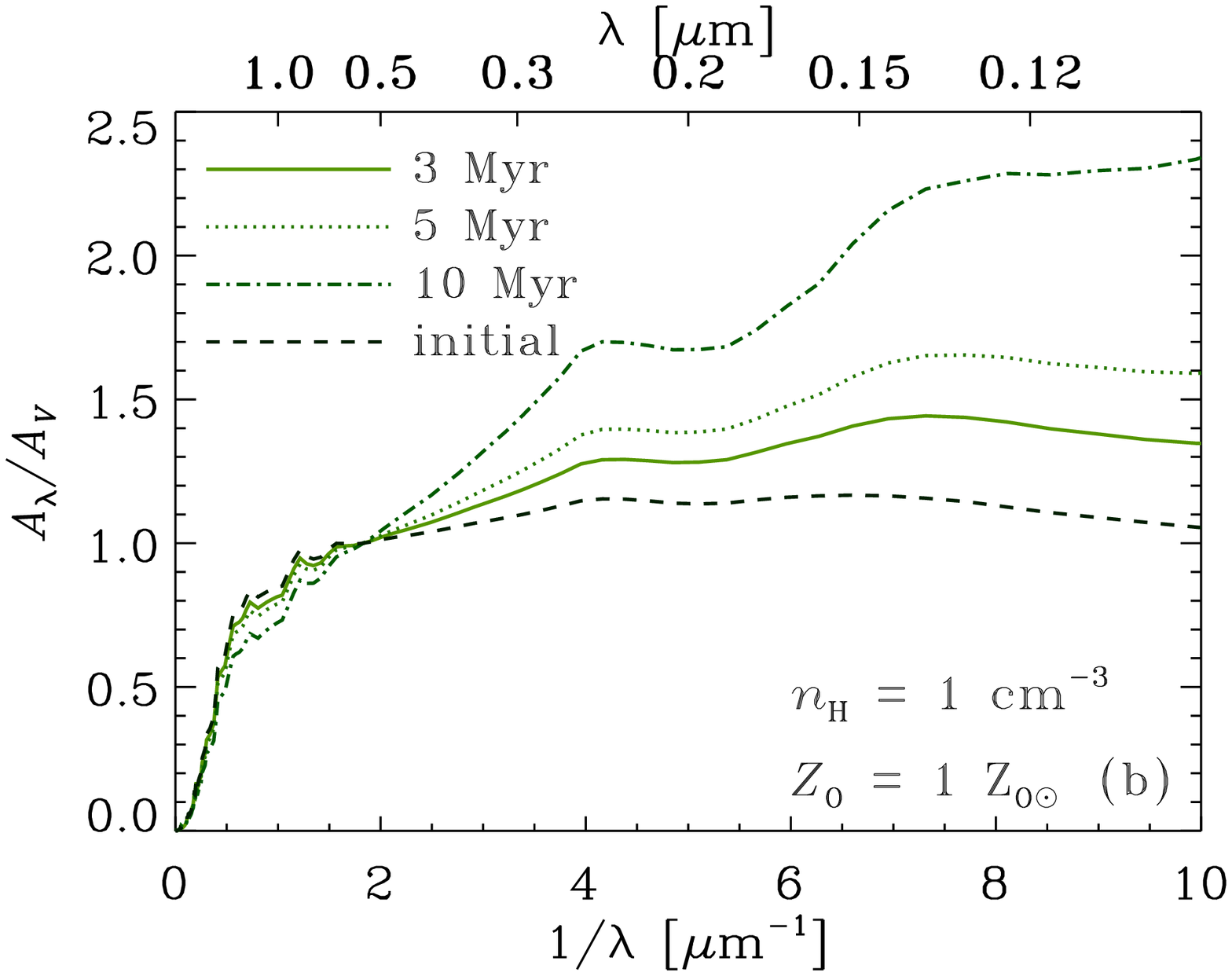}
\includegraphics[width=0.45\textwidth]{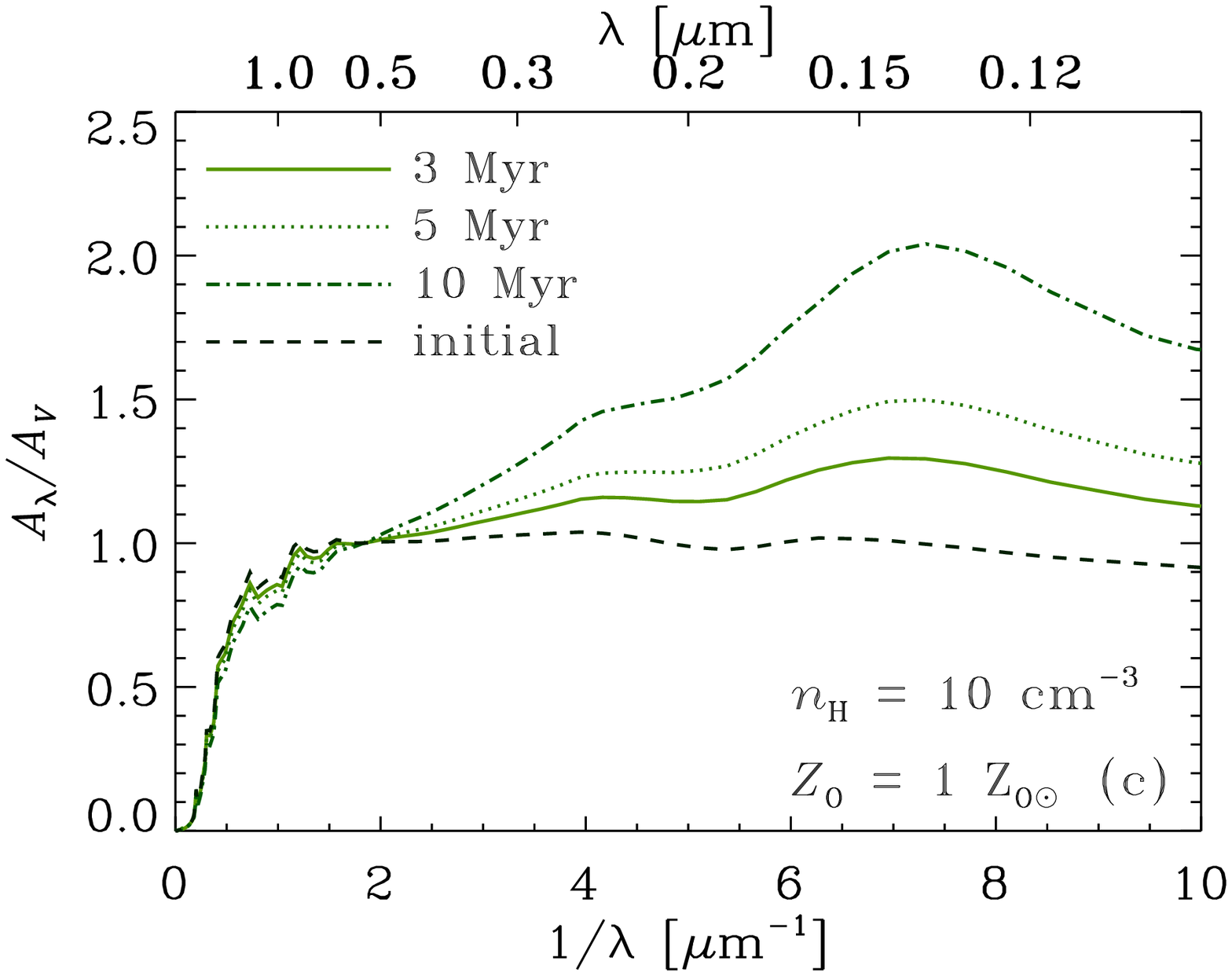}
\includegraphics[width=0.45\textwidth]{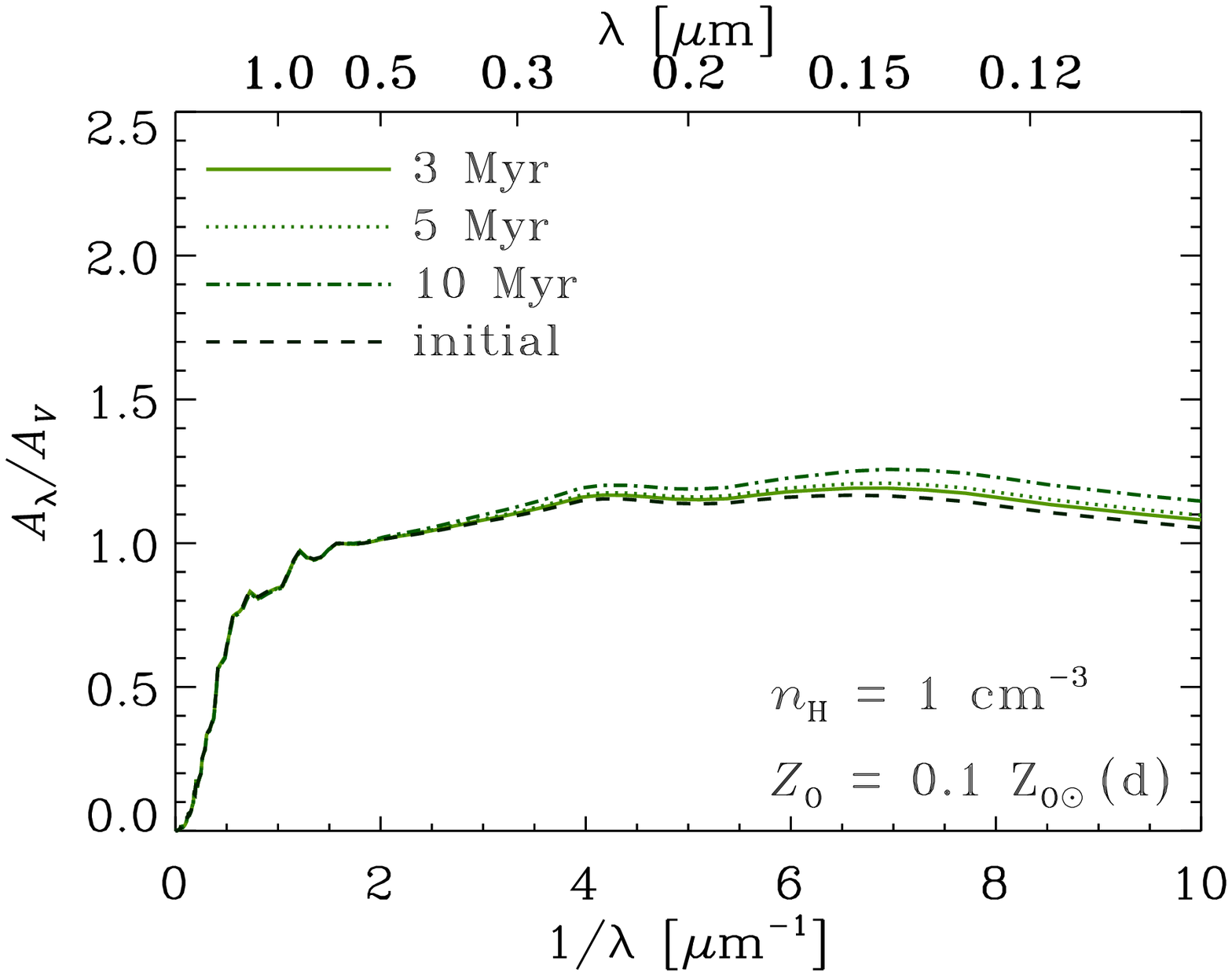}
 \caption{Extinction curves normalized to the $V$ band
extinction. The solid, dotted, and dot-dashed lines
indicate the results at $t=3$, 5, and 10 Myr,
respectively. Panels (a), (b), and (c)
present the results for $\nH =0.1$, 1, and 10 cm$^{-3}$,
respectively, with $\ZO=1~\ZOsun$.
Panel (d) shows the result for $\nH =1~\mathrm{cm}^{-3}$
with $\ZO=0.1~\ZOsun$.}
 \label{fig:ext_age}
\end{figure*}

In Fig.\ \ref{fig:ext_age}, we show the time variation of
extinction curves for $\nH =0.1$, 1, and 10 cm$^{-3}$
with $\ZO =1~\ZOsun$. We normalize the extinction
to $A_V$ (i.e.\ at $\lambda =0.55~\micron$). As stated
by \citet{hirashita08}, the initial extinction curve is
steeper for lower $\nH$, since more small grains
survive after {the shock destruction in SNRs}.
We also observe
that the extinction curve becomes steeper as the
grains are shattered for a longer
time because of the production of small grains. Indeed,
at $t=5$ Myr, $A_\lambda /A_V$ at
$\lambda\sim 0.2~\micron$
increases by more than 20\% for $\nH =1$ cm$^{-3}$.
The variation of
the slope by shattering is more pronounced for larger
$\nH$ since the original extinction curve is flatter.

\begin{figure}
\includegraphics[width=0.45\textwidth]{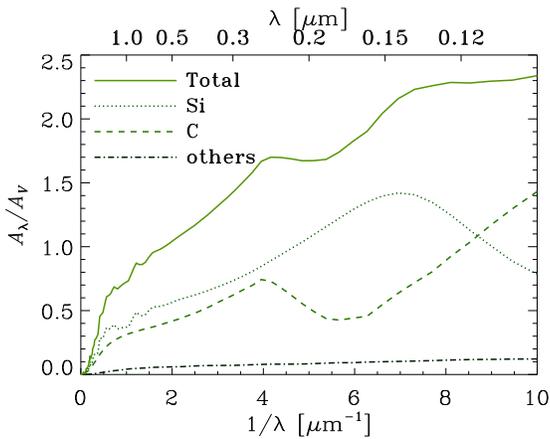}
 \caption{Contributions from Si, C, and the other grain
species (dotted, dashed, and dot-dashed lines,
respectively) for the case of Fig.\ \ref{fig:ext_age}b
($\nH =1~\mathrm{cm}^{-3}$ and $\ZO=1~\ZOsun$) at
$t=10$ Myr.
The solid line shows the total extinction.}
 \label{fig:ext_component}
\end{figure}

The extinction curves are dominated by Si and C, which
survive after the shock destruction in SNRs because of
the relatively large sizes (N07). Therefore, the
steepening of extinction curve is mainly due to the
production of small Si and C grains by shattering.
The contributions from Si, C, and the other species
are shown in Fig.~\ref{fig:ext_component}. The `bump'
features at $1/\lambda\sim 4$ and
7 $\micron^{-1}$ originate from the absorption by
C and Si, respectively. Such features tend to be
prominent for smaller grains, {since as grains
become larger the extinction cross sections are more
determined by the geometrical ones, not by the
grain properties \citep{bohren83}.}
Thus, not only the steep slope but also various
features in the extinction curve become apparent
as grains suffer from shattering.

We also examine the dependence on the dust abundance
(metallicity). In Fig.\ \ref{fig:ext_age}d, we show
the evolution of extinction curve for
$\ZO =0.1~\ZOsun$. The effect of shattering
is significantly reduced compared with the solar
metallicity case. If the grain velocity as a function
of grain size is fixed, a constant $\ZO t$ gives the
same result. Thus, if $\ZO$ is ten times lower,
10 times longer time is required for the same shattering
effect to appear. The scaling with $\ZO t$ is useful if
one would like to know the results for other
time-scales and/or metallicities.

\section{DISCUSSION}\label{sec:discussion}

\subsection{Steepening of UV extinction curve}
\label{subsec:steep}

In the above section, we have shown that
{the abundance of small ($a<0.1~\micron$)
grains increase by shattering of large
($a\ga 0.1~\micron$) grains. Consequently,
the slope of the UV
extinction curve becomes steep after shattering.}
Here, we discuss this issue in terms
of the grain size distribution.

The contribution from grains in a logarithmic size
range [$\ln a$, $\ln a +\mathrm{d}\ln a$] to the
extinction can be
written as $\mathrm{d}\kappa_\mathrm{ext}\equiv
\pi a^2Q_\lambda (a)n(a)a\,\mathrm{d}\ln a$,
where $Q_\lambda (a)$ is the extinction cross
section normalized to the geometrical cross section
($\pi a^2$). If the size distribution is approximated
by a power-law ($n\propto a^{-p}$) over a certain size
range,
$\mathrm{d}\kappa_\mathrm{ext}/\mathrm{d}\ln a\propto
a^{3-p}Q_\lambda(a)$.
Since $Q_\lambda (a)\sim 1$ for $2\pi a\ga\lambda$ and
$Q_\lambda (a)\propto a$ for $2\pi a\ll\lambda$
\citep[e.g.][]{bohren83}, we obtain
$\mathrm{d}\kappa_\mathrm{ext}/\mathrm{d}\ln a\propto a^{3-p}$
for $2\pi a\ga\lambda$ and
$\mathrm{d}\kappa_\mathrm{ext}/\mathrm{d}\ln a\propto a^{4-p}$
for $2\pi a\ll\lambda$. Thus, if $p<3$, the largest
grains have
the largest contribution to the extinction. In order for
small grains to
have significant contribution to the extinction,
$p\geq 3$ should be satisfied. If $3<p<4$, the largest
contribution to the extinction comes from the grains
with $2\pi a\sim\lambda$. In other words, the UV
($\lambda\sim 0.2~\micron$) extinction curve is steepened
significantly if grains with $a\sim 0.03~\micron$ are
produced and $p\ga 3$ is satisfied around this grain size.

{}From Figs.\ \ref{fig:size_n0.1}--\ref{fig:size_n10},
we observe that a large number of grains with
$a\sim 0.03~\micron$ are produced and the slope around
this grain radius is $p\ga 3$ for the solar metallicity
cases. Indeed, the UV slope of extinction
curve is steepened for the solar metallicity cases as
shown in Fig.\ \ref{fig:ext_age}.

In this paper, the shattered fragments are distributed
with a size distribution with exponent
$\alpha_\mathrm{f}=3.3$ (Section \ref{subsec:shatter}).
\citet{jones96}, from a discussion on the cratering flow,
argue that the value of
$\alpha_\mathrm{f}$ slightly larger than 3 is robust.
Even if $\alpha_\mathrm{f}=2.5$ is assumed as an extreme
case, the difference
in the extinction curve is less than 10\% at
$\lambda =0.1~\micron$ and smaller at longer wavelengths
(see the Appendix \ref{app:alpha2.5} for details).

\subsection{Grain properties in starburst environments}

{}From the results above, the presence of small grains
in starburst environments is generally predicted,
although SNe II tend to eject large grains because of
{the shock destruction in SNRs}. For example, BCDs
(or H \textsc{ii} galaxies) in the nearby Universe
host large ionized region and the age of the current
star formation episode is a few~Myr--20 Myr
\citep[e.g.][]{hirashita04,takeuchi05}. These ages are
just in the range where shattering could modify the
grain size distribution and extinction curve, although
we should take into account the low metallicity in
BCDs. Some BCDs show an excess of near-infrared
emission \citep[e.g.][]{hunt01}, which can be attributed
by the emission from transiently
heated very small grains
\citep{aannestad79,sellgren84,draine85}.
\citet{galliano05} have carried out a comprehensive
analysis of the SEDs of dust and stars
in some dwarf galaxies (dwarf irregular galaxies and BCDs),
and have shown that the grain size is biased to small grains
with $\sim$ a few nm. Since their sample galaxies have
metallicities larger than 1/10 Z$_{\sun}$, shattering in WIM
can work as a production source of nm-sized grains
on time-scales of a few Myr and thus can be considered as
an origin of small grains in these galaxies.

It is natural to expect that a similar condition (i.e.\
turbulence in WIM sustained more than a few Myr) is
generally realized in starburst galaxies. Although it
is hard to
compare the extinction curve with the observed
wavelength dependence of the dust attenuation
because of the effects of radiative transfer
\citep{calzetti01,inoue05}, shattering may be crucial
to reproduce the reddening in starburst galaxies.
Therefore, shattering should be considered as a
source of small grains, which contribute to
the reddening.
{Or dust produced by AGB stars in the
underlying old
population (older than several $\times 10^8$ yr;
\citealt{valiante09}) could contribute to the steepening
if they produce small grains; however, there are
some observational indications that dust grains
produced in AGB stars are large ($a\sim 0.1~\micron$)
\citep{groenewegen97,gauger99}.}

Efficient shattering also occurs in the ISM by the
passage of SN shocks. \citet{jones96} show that a
large fraction of large
grains with $a>0.1~\micron$ is redistributed into
smaller grains by a single passage of shock with a
velocity of $\sim 100$ km s$^{-1}$.
{In their calculation,
large grains take longer time before they are
dynamically coupled
with gas}
and are subject to more collisions with dust.
\citet{jones96}
consider the MRN distribution as the initial grain size
distribution, which enhances the shattering efficiency
compared with our case,
because of the enhanced collision with the abundant
small grains. Below, we estimate the time-scale on which
shattering in SN shocks destroys large grains based
on \citet{jones96}, although the time-scale obtained might
be an underestimate {for the grains produced by SNe II,
because of the enhanced collision rate in the MRN
distribution}.

The time-scale on which shattering in SN shocks effectively
destroys large grains can basically estimated by
a similar way to \citet{mckee89}.
A single SN can sweep
$M_\mathrm{sw}\sim 10^4~\Msun$ of gas
(i.e.\ $M_\mathrm{sw}v_\mathrm{s}^2/2\sim E_\mathrm{SN}$
with a shock velocity $v_\mathrm{s}\sim 100$ km s$^{-1}$
and energy
given to gas by a SN
$E_\mathrm{SN}\sim 10^{51}$ erg).
Then, the gas mass
swept by SN shocks with
$v_\mathrm{s}\ga 100$ km s$^{-1}$ per unit time
can be estimated as
$M_\mathrm{sw}\gamma$, where $\gamma$ is
the SN rate. Thus, the time-scale on which the
entire gas mass $M_\mathrm{g}$ is affected by
shattering by SN shocks is estimated as
$\tau_\mathrm{sw}\sim M_\mathrm{g}/
(M_\mathrm{sw}\gamma)$.
Since
$\gamma/\psi\sim 10^{-2}\Msun^{-1}$ for a Salpeter
initial mass function \citep{salpeter55}
($\psi$ is the star formation rate), the above
time-scale is estimated as
$\tau_\mathrm{sw}\sim 10^{-2}M_\mathrm{g}/\psi$.
This estimate indicates that the shattering time-scale
by SN
shocks is about 0.01 times the gas consumption
time-scale by star formation. In starburst environments,
$M_\mathrm{g}/\psi\sim 10^8$--$10^9$ yr may be
reasonable \citep{young86}, and shattering in SN shocks
occurs in 1--10 Myr, which is comparable to the
time-scale investigated in this paper. Therefore,
both shattering in turbulence and that in SN
shocks can affect the grain size distribution.
A detailed calculation of shattering in SN
shocks of grains produced by SNe II is required
before we judge which of these two shattering
mechanisms is dominated.

It might be also useful to discuss our results in terms
of the extinction curves of the Large and Small
Magellanic Cloud (LMC and SMC), both of which have
developed H \textsc{ii} regions such as 30~Doradus.
Indeed, \citet{bernard08} indicate that the
70 $\micron$ excess around 30~Doradus can be explained
by an enhancement of the
abundance of very small grains possibly by the
destruction of large grains. \citet{bot04} find this excess
in the SMC. \citet{paradis09} show that
the very small grain abundance is really enhanced
around 30 Doradus by using an SED model of dust
emission. However, the extinction curves
in these galaxies are much steeper than our results
($A_\lambda /A_V\simeq 2.9$ and 3.2 at
$\lambda\simeq 0.2~\micron$ for the LMC and the
SMC, respectively; \citealt{pei92}). Since those
galaxies have
less intense star formation than BCDs, it is hard
to extract the starbursting components where
shattering of large grains should be
working as investigated in this paper. The steep
extinction curves of the LMC
and the SMC indicate that we should consider not only
the dust production/shattering in star-forming
regions but also some other mechanisms which
act as efficient production sources of small grains.
For example, shattering
in warm \textit{neutral} medium works on a time-scale
of 100 Myr (HY09). ISM phase exchange, which
occurs on a time-scale of 50--100 Myr, also affects
the evolution of grain size distribution
\citep{odonnell97}.
Such longer-time-scale mechanisms could also have
affected the extinction
curves (grain size distributions) of those galaxies.
The current paper, which
focuses on a short-time-scale ($<10$ Myr)
grain processing, is a starting point to
include other physical
processes in future work.

\subsection{Comparison with high-$z$ data}
\label{subsec:comp}

At $z>5$, it is usually assumed that the main
production source of dust is SNe II whose
progenitors have short
lifetimes, since the cosmic age is too young
for low mass stars to evolve \citep[but see][]{valiante09}.
Thus, the extinction curves at such high $z$ are
often used to test the theory of dust production in
SNe II \citep{maiolino04,hirashita05}. As a representative
case of observed high-$z$ extinction curve, we discuss
the restframe UV extinction curve of
\sdss\ ($z=6.2$) obtained by \citet{maiolino04}.

\begin{figure}
\includegraphics[width=0.45\textwidth]{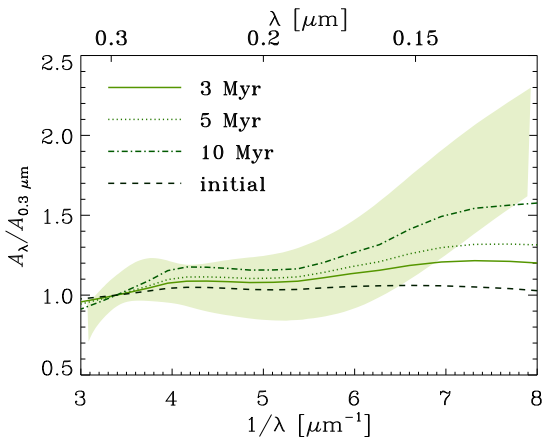}
 \caption{The same extinction curves as those shown in
Fig.\ \ref{fig:ext_age}b are plotted only in the UV
range ($\ZO =1~\ZOsun$ and $\nH =1$ cm$^{-3}$).
The shaded area shows the observed extinction curve
for \sdss\ ($z=6.2$) by \citet{maiolino04}, including
the uncertainty.}
 \label{fig:ext_uv}
\end{figure}

In Fig.\ \ref{fig:ext_uv}, we show the UV part of the
extinction curves calculated by our models in
comparison with the observed UV extinction curve
of \sdss. The extinction curves are normalized
to the value at $\lambda =0.3~\micron$. We show the
result for $\nH =1~\mathrm{cm}^{-3}$,
{but the following
discussions hold qualitatively also for other densities.}
As discussed in
\citet{hirashita08}, the initial extinction curve
before shattering is too flat to explain the UV rise
in the observed
extinction curve because small grains are selectively
destroyed in SNRs. However, after shattering,
the extinction curve approaches the observed curve
because of the production of small grains. After
10 Myr of shattering, the observed
extinction curve is reproduced. Not only the slope
but also the bump feature at
$1/\lambda\sim 4~\micron^{-1}$, which becomes
prominent after
shattering {(Section \ref{subsec:theor_extinc})},
may account for the behaviour of the
observed extinction curve around
$1/\lambda\sim 3.5$--4 $\micron^{-1}$.

In summary, if the metallicity is nearly solar and the age
of the current episode of starburst is larger than 5 Myr, we
should take the effect of shattering in turbulence into
account in comparing the observed extinction curve with the
theoretical one even at $z>5$. Since quasars tend to be
found in evolved stellar system whose metallicity could be
nearly solar (or more than solar; \citealt{juarez09}), the
UV rise of the extinction curve may be
caused by the production of small grains by shattering.
The dependence of the extinction curve on age and
metallicity may also be responsible
for the variation of UV slope of
the quasar spectra in the sample of
\citet{maiolino04a}.

\subsection{Remarks on grain physics}
\label{subsec:remark}

{
Before concluding this paper, we mention some physical
processes to be considered in the future. In the
calculation of {the shock destruction of grains in SNRs}
by N07, the effects of grain electrical charge and the
effects of magnetic fields are ignored. As shown in
\citet{jones94,jones96} and more recently by
\citet{guillet07} and \citet{guillet09}, the dynamics of
charged grains is critically modified by magnetic fields.
The gyration around the magnetic fields tend to
strengthen the coupling between gas and dust, and this
effect could suppress the ejection of large grains
into the ISM. Thus, not only small grains but also
large grains with $a\ga 0.1~\micron$ could be subject
to significant processing in the shock.
\citet{slavin04} show that the presence of magnetic
fields in shocks produce complexity in the kinematics of
large ($\ga 0.1~\micron$) grains. Thus,
it may be important to trace the grain trajectory
around the reverse and forward shocks.
The quantification of all these effects of magnetic fields
is left for future work.}

{
Nevertheless, the importance of shattering by turbulence 
for small-grain production in starburst galaxies should be
an  important issue even if we consider the the effect of
magnetic  field in the future, because it is still true that
the shock destruction
in SNRs  suppresses the injection of small grains into the
ISM. It should also be kept in mind that at the smallest size
ranges (a few \AA), the treatment of grains as bulk solid
may not be a good approximation. Since such tiny grains
do not affect the UV--optical extinction curve
as discussed in Section \ref{subsec:steep},
the results on the extinction curves are not
affected. Mid-infrared spectra of dust emission
are more suitable
to constrain the abundance of such small grains
\citep[e.g.][]{mathis90}.
}

\section{Conclusion}\label{sec:conclusion}

We have theoretically investigated the effect of
shattering in {turbulent WIM on the grain size
distribution by using the framework for shattering by
\citet{jones94,jones96} and the calculation
of interstellar MHD turbulence by \citet{yan04}.}
We have focused on
systems in which dust is predominantly produced
by SNe II. Although SNe II tend to eject large
($a\ga 0.1~\micron$) grains because of {the shock
destruction in SNRs} (N07), shattering in WIM supplies
small grains on a time-scale of several Myr
{in the
solar-metallicity (i.e.\ Galactic dust-to-gas ratio)
case}. Consequently,
the extinction curve is steepened and the features such
as the carbon bump around
$1/\lambda\sim 4~\micron^{-1}$
and the Si bump around $1/\lambda\sim 7~\micron^{-1}$
become apparent if the metallicity is solar and the
duration of shattering is longer than $\sim 5$ Myr.
Therefore, when we treat a system in
which the metallicity is solar and the star formation age is
$\ga 5$ Myr, we should take into account the
effect of shattering in interstellar turbulence.
In particular, the
extinction curves of high-$z$ quasars, whose metallicity
is typically (above) solar, may be affected
by shattering, and the UV rise of the extinction curve
as well as the bump feature at
$1/\lambda\sim 3.5$--4 $\micron^{-1}$ can be
attributed to the small grains produced by shattering.
{If the metallicity is $\la 1/10$ solar, the
extinction curve
does not vary significantly on a time-scale of $\la 10$ Myr
because the frequency of grain--grain collision is
reduced in proportion to the grain abundance. Thus, the
steepening mechanism of extinction curve discussed in
this paper
is valid for systems whose metallicities are significantly
larger than 1/10 solar.}
We conclude that shattering in WIM is generally of potential
importance in starburst galaxies as a production mechanism
of small grains.

\section*{Acknowledgments}
We thank the referee, A. P. Jones, for useful comments 
which improved this paper considerably. We thank
T. T. Takeuchi and T. T. Ishii for helpful discussions.
HY is supported by the TAP fellowship in Arizona.
TN has been supported by World Premier International
Research Center Initiative (WPI Initiative), MEXT, Japan,
and by the Grant-in-Aid for Scientific Research of the
Japan Society for the Promotion of Science (19740094,
20340038).

\appendix

\section{Test for the `one-species' method}
\label{app:individual}

As stated in Section \ref{subsec:shatter}, all the grain
species other than carbonaceous grains are treated as
a single species, called formally `silicate' in calculating
the grain size distribution. This approximation is called
`one-species' method, and it is exact if all the grain
species have the same shape of
grain size distribution. We expect that the one-species
method gives a reasonable answer since Si is dominated
among the `silicate' category.
Although the `silicate' species other than
Si (we call these species non-Si grains) have minor
contributions in grain mass, some of them have
a significant contribution to the number of small-sized
grains, which affect the UV slope of the extinction curve.
Here we
test the validity of the one-species method in
comparison with
the `individual-species method' as explained below.

\begin{figure*}
\includegraphics[width=0.45\textwidth]{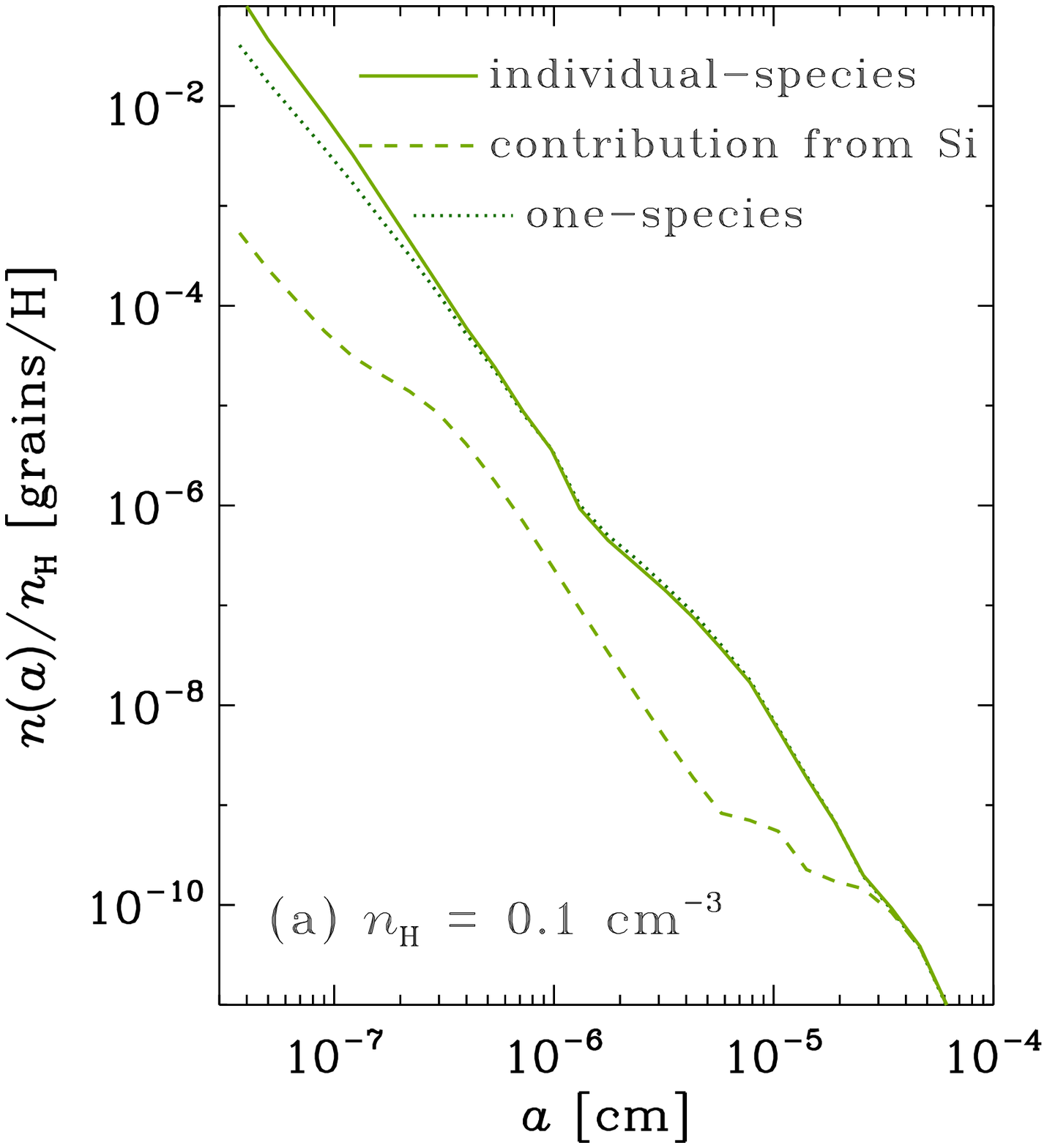}
\includegraphics[width=0.45\textwidth]{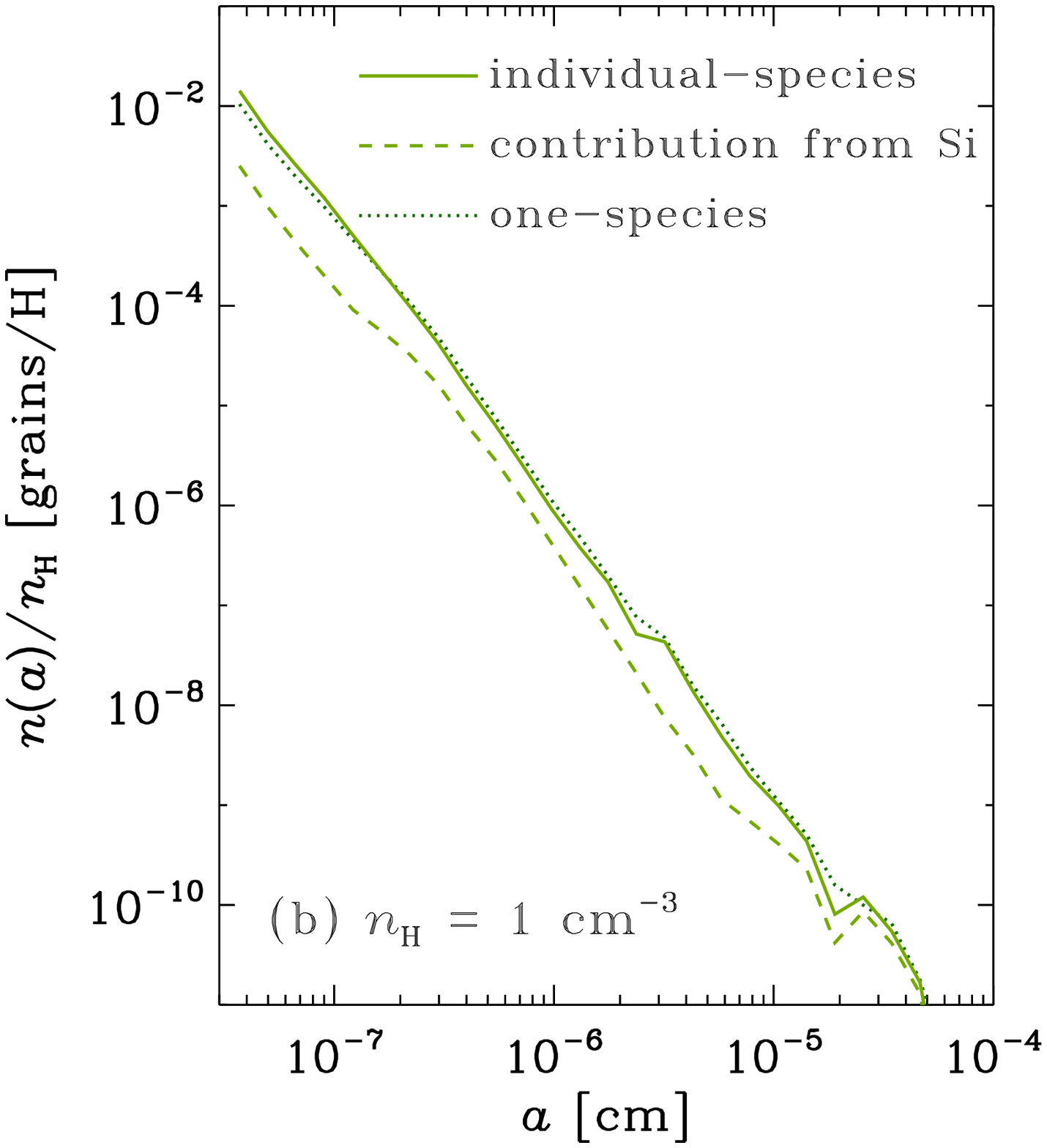}
 \caption{Size distributions for (a) $\nH =0.1$ cm$^{-3}$ and
(b) $\nH =1$ cm$^{-3}$ for the grains other than carbon
(i.e.\ `silicate'). The solid and dotted lines show
the results with the individual-species method and with
the one-species method, respectively. The dashed line
presents the contribution from Si to the solid line. The metallicity
and the age are assumed to be 1 $\ZOsun$ and 5 Myr,
respectively.}
 \label{fig:size_individual}
\end{figure*}

\begin{figure*}
\includegraphics[width=0.45\textwidth]{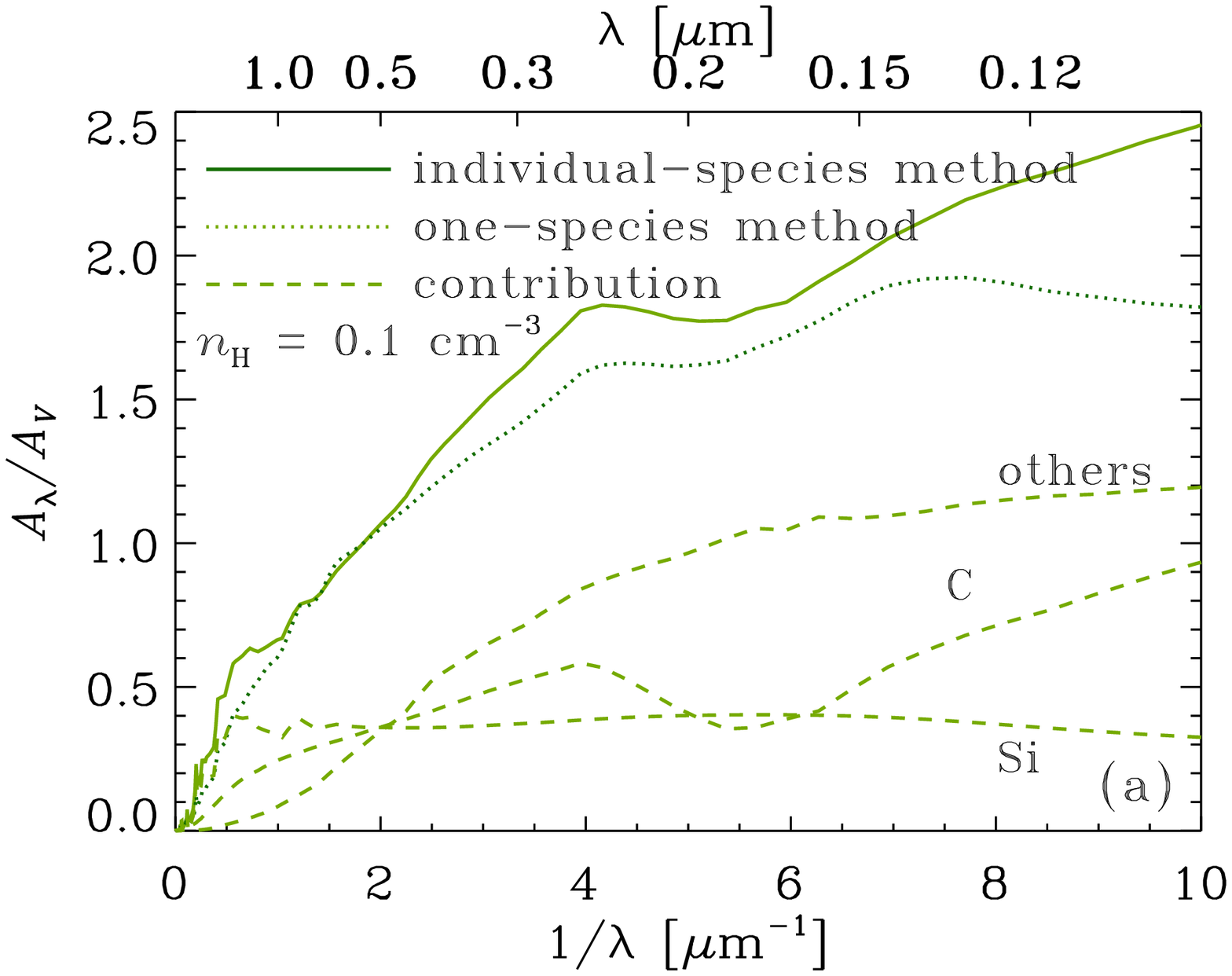}
\includegraphics[width=0.45\textwidth]{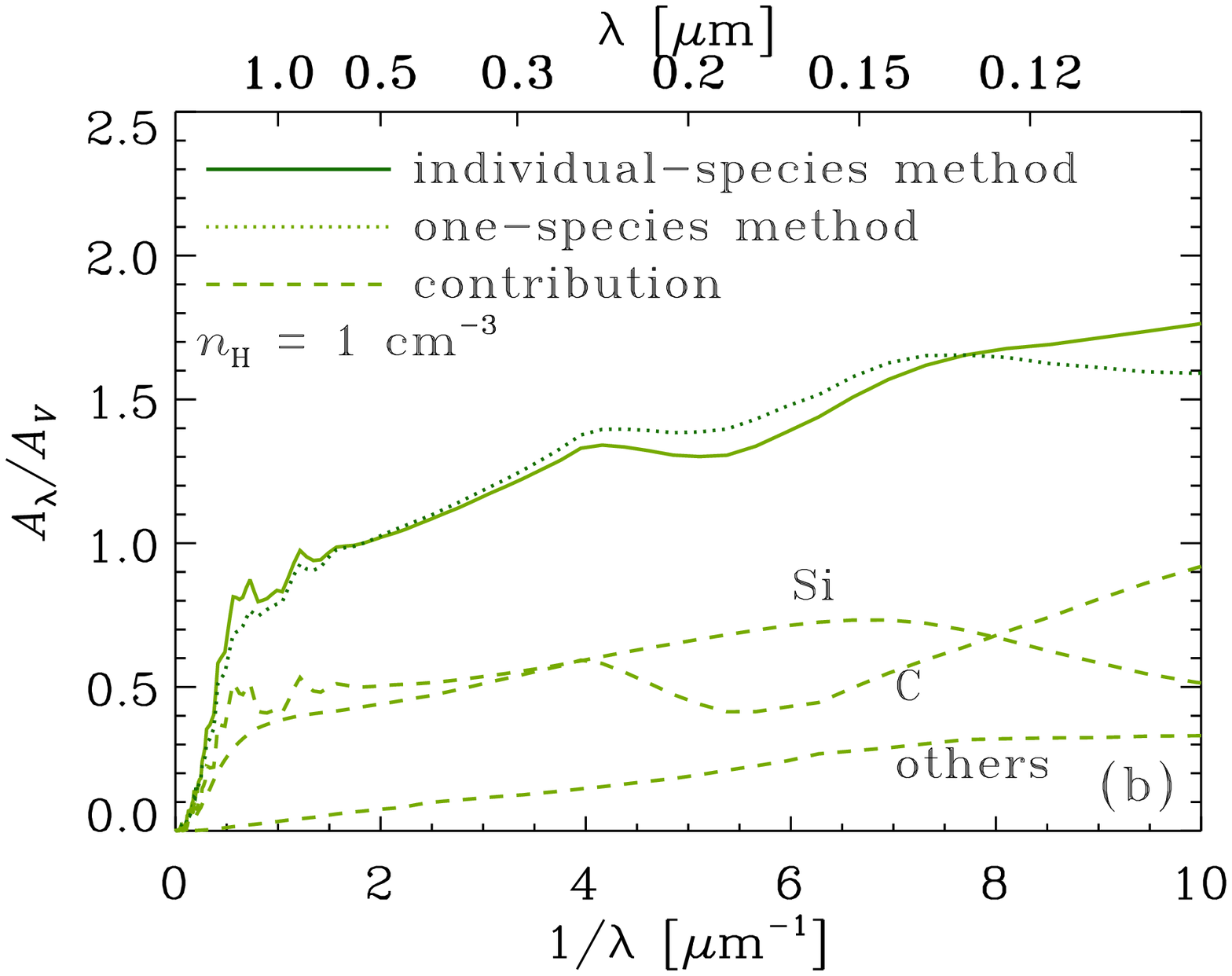}
 \caption{Extinction curves normalized to the $V$ band
extinction for the grain size distributions presented in
Fig.\ \ref{fig:size_individual} (the grain size distributions of
carbonaceous grains are the same as those in
Figs.\ \ref{fig:size_n0.1} and \ref{fig:size_n1}).
Panels (a) and (b) present the
cases with $\nH =0.1$ and 1 cm$^{-3}$, respectively.
The solid and dotted lines show the results of the
individual-species and one-species methods, respectively.
The dashed
line represents the contributions from various species
(C, Si, and the others as labelled in the figures) for the
individual-species method.}
 \label{fig:ext_individual}
\end{figure*}

The `individual-species' method adopts the grain size
distribution of individual species and the evolution
of grain size distribution is separately calculated for
individual species (note that the grain size distribution
summed over all the species other than carbonaceous
grains is adopted for `silicate' in the one-species
method). In calculating the evolution of grain size
distribution of a certain species, the total mass
density of the
species relative to the gas density is assumed to be
the total dust-to-gas ratio (but the grain size distribution
after shattering is normalized again
to recover the correct mass ratio of each species).
This treatment maximizes the
production of small grains for non-Si species,
which have smaller sizes than Si,
but minimizes the production of small Si grains.
Thus, this method is suitable to examine the maximum
possible contribution
from non-Si small grains to the UV extinction curve.

In Fig.\ \ref{fig:size_individual}, we compare the grain
size distributions predicted by the one-species and
individual-species methods for $\nH =0.1$ and
1 cm$^{-3}$ at 5 Myr. For $\nH =10$ cm$^{-3}$, the
difference between the two methods is
negligible because non-Si grains contribute little
to the total grain abundance. From the figure, we
observe that the
difference is
relatively large in the case of $\nH =0.1$ cm$^{-3}$.
This is because the fraction of non-Si grains is larger
for $\nH =0.1$ cm$^{-3}$ than for
$\nH =1$ cm$^{-3}$.

In Fig.\ \ref{fig:ext_individual}, we show the extinction
curves calculated by the two methods. We observe
that the extinction curves of the individual-species
method tend to be steeper than those of the one-species
method. As can be seen in the figure, the steeper slope
comes from the contribution from the non-Si grains
indicated by `others'. In the individual-species method,
the size distribution of each non-Si species, which has
a larger fraction of small grains than that of Si, is
calculated separately, so that the production of small
non-Si grains is enhanced.
We note that the `real' grain size distribution would lie
between the results of the two methods. This means the
approximate
treatment adopted in the text (i.e.\ one-species method)
is justified for $\nH\ga 1$ cm$^{-3}$. For
$\nH\la 0.1$ cm$^{-3}$, because the contribution from
non-Si species is significant, the error of the
one-species method is at most $\sim 10$\% at
$\lambda\sim 0.2~\micron$, and $\sim 40$\% at
$\lambda\sim 0.1~\micron$. In order to overcome this
uncertainty, we should develop a different scheme that
could treat the collisions between multiple species
(in our case, 9 species),
which the current scheme cannot treat in a reasonable
computational time.

\section{Fragment size distribution with a shallower
slope}\label{app:alpha2.5}

\begin{figure*}
\includegraphics[width=0.45\textwidth]{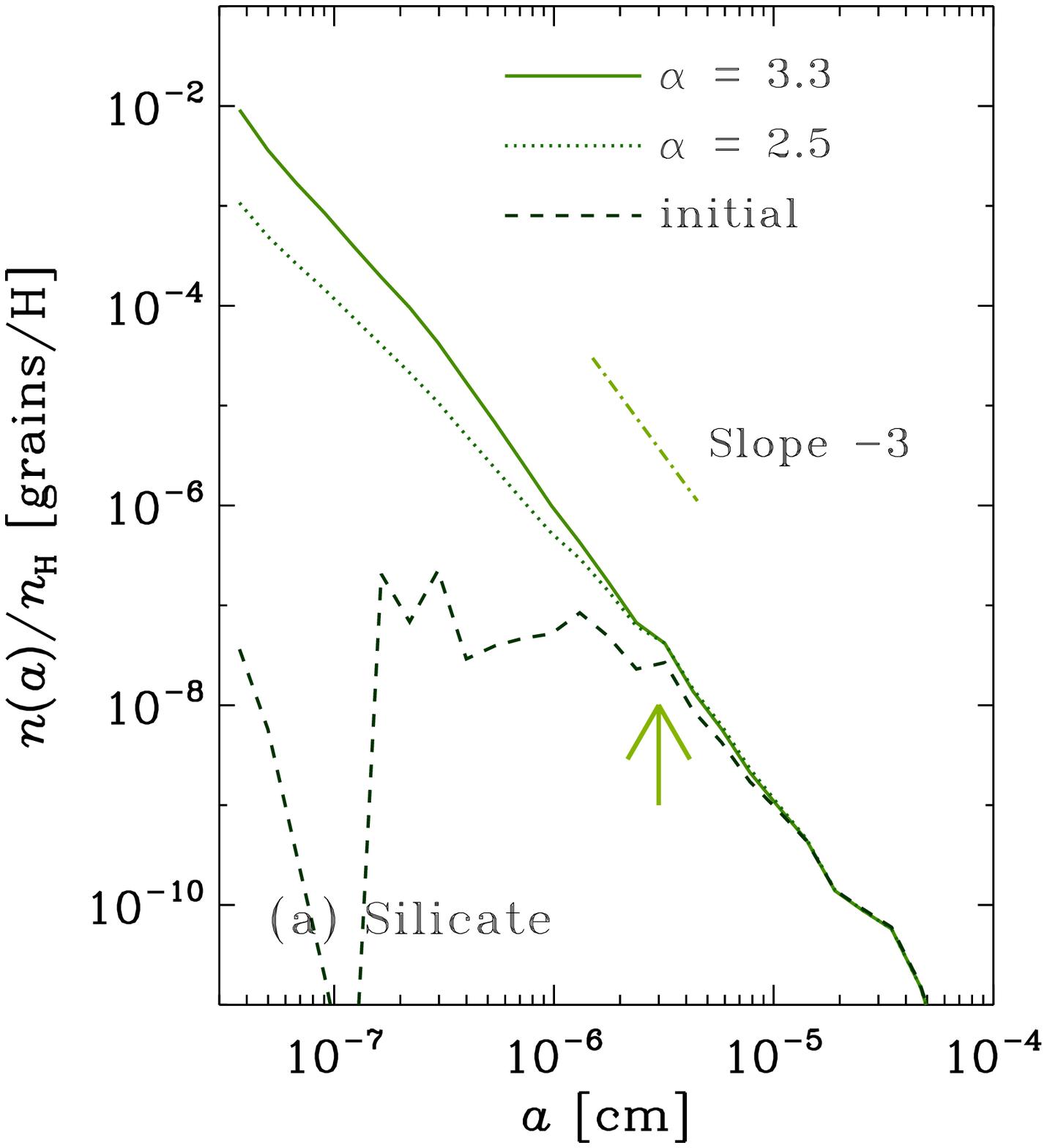}
\includegraphics[width=0.45\textwidth]{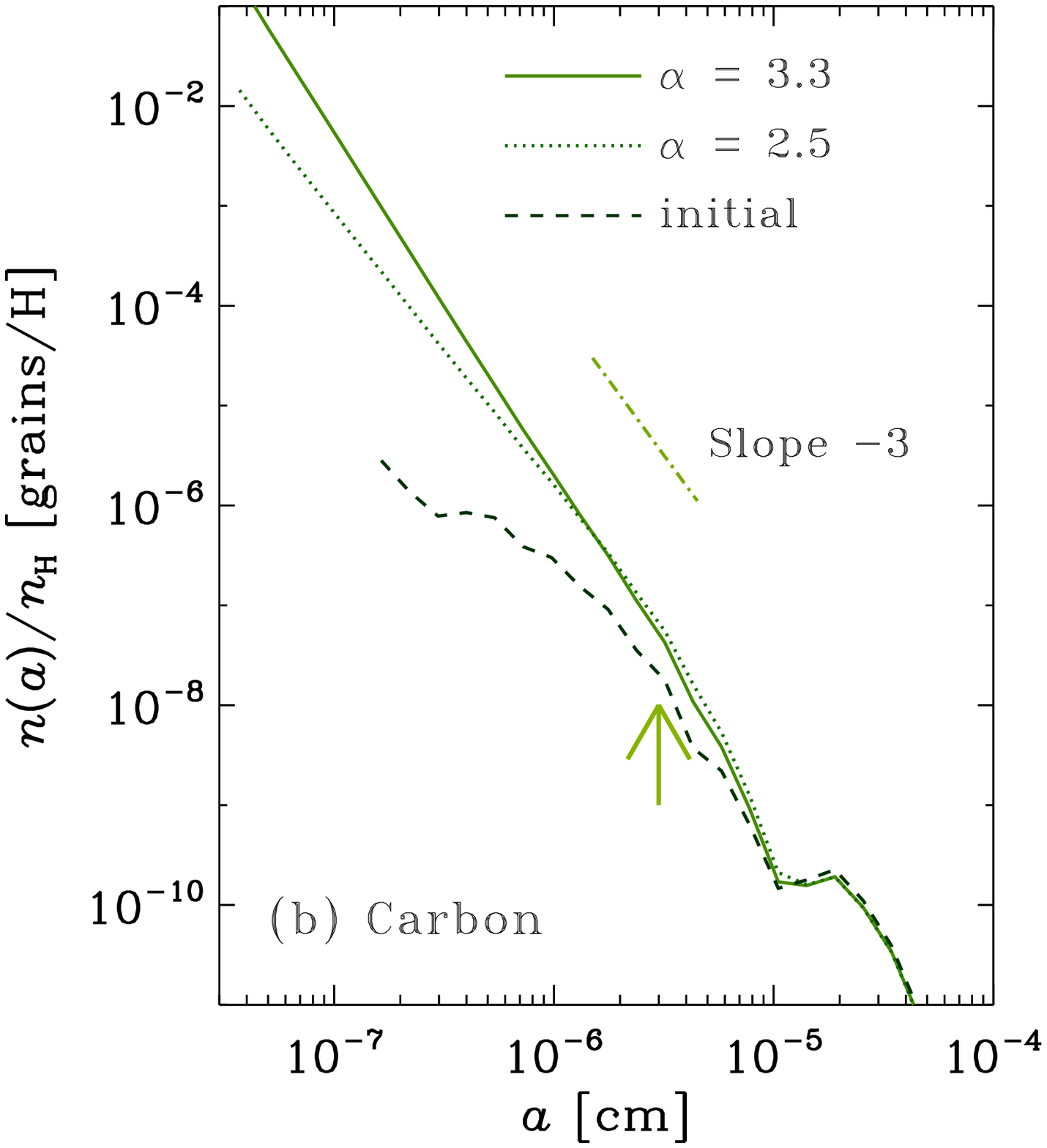}
 \caption{Grain size distributions for $\nH =1$ cm$^{-3}$
with a metallicity of 1 $\ZOsun$. The solid and dotted lines
show the results at $t=5$ Myr for $\alpha_\mathrm{f}=3.3$
and 2.5, respectively. The dashed line presents the initial
grain size distribution before shattering.
Two grain species,
(a) silicate and (b) carbonaceous dust, are shown.
{The arrow is put at
$a=0.03~\micron$ as a rough representative size of the
grains contributing to the steepening of the UV
extinction curve.}}
 \label{fig:size_alpha}
\end{figure*}

The size distribution of shattered fragments is
assumed to be a power law with an exponent of
$-\alpha_\mathrm{f}$. As we discuss in the text, the
steepening of extinction curve becomes prominent if
the power-law exponent ($p$) of the grain size
distribution around $a\sim 0.03~\micron$
is steeper than $\sim 3$ (Section \ref{subsec:steep}).
\citet{jones96} have shown
that the size distribution after shattering is not
sensitive to $\alpha_\mathrm{f}$. They also argue that
$\alpha_\mathrm{f}$ slightly larger than 3 is robust
against the change of the cratering flow parameters in
shattering ($\alpha_\mathrm{f}=3.3$ is adopted
in the text). Nevertheless it would be interesting
to examine if $p>3$ is realized even if we assume
$\alpha_\mathrm{f}<3$.

\begin{figure}
\includegraphics[width=0.45\textwidth]{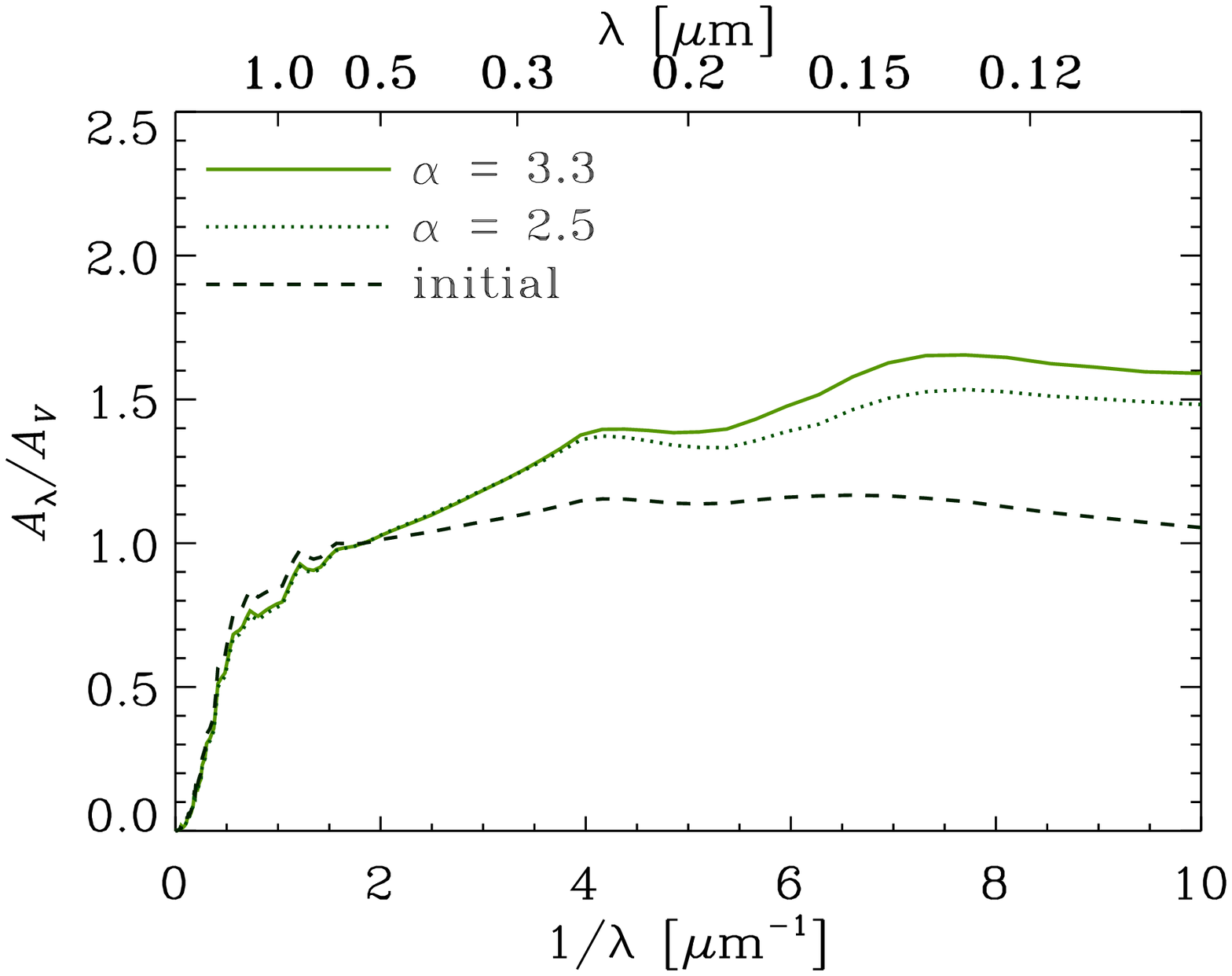}
 \caption{Extinction curves normalized to the $V$ band
extinction for the grain size distributions presented in
Fig.\ \ref{fig:size_alpha}.
The solid and dotted lines show the results for
$\alpha_\mathrm{f}=3.3$ and 2.5, respectively. The dashed
line represents the initial extinction curve before
shattering.}
 \label{fig:ext_alpha}
\end{figure}

Here we examine the smallest exponent adopted in
\citet{jones96}, $\alpha_\mathrm{f}=2.5$ as an
extreme case. The ambient hydrogen number density is
fixed to $\nH =1~\mathrm{cm}^{-3}$.
In Fig.\ \ref{fig:size_alpha}, we show the result at
$t=5$ Myr. As expected, the effect of $\alpha_\mathrm{f}$
is more prominent for smaller grains, since shattering
with large $\alpha_\mathrm{f}$ can supply small grains
more efficiently. However, we
observe that the difference between
$\alpha_\mathrm{f}=2.5$ and 3.3 is small around
$a\sim 0.03~\micron$, confirming the result
of \citet{jones96}. The small difference comes from
the fixed shattered mass in a collision; that is,
the distribution of
grain fragments as a function of size has a minor
effect compared with the total mass of shattered
fragments (shattering efficiency).

The extinction curves are shown in
Fig.\ \ref{fig:ext_alpha}. We observe that the
difference between the two curves with $\alpha =2.5$
and 3.3 is negligibly small at
$\lambda\sim 0.3~\micron$ and is less than 10\% even at
$\lambda\sim 0.1~\micron$. The small difference is the
natural consequence of the small variation of grain size
distribution at $a\ga 0.03~\micron$.

\bsp

\label{lastpage}

\end{document}